\documentclass[]{spie}  

\usepackage{amsmath,amsfonts,amssymb}
\usepackage{graphicx}
\usepackage[dvipsnames]{xcolor}
\usepackage{subcaption,booktabs}
\usepackage{subcaption}
\usepackage[colorlinks=true, allcolors=blue]{hyperref}
\usepackage{tabularx}
\title{Speckle simulation tool for automated modelling of a large range of telescope aperture to fried parameter ratios}

\author[a]{Sorabh Chhabra}
\author[a]{Abhay A. Kohok}
\author[a]{Bhushan S. Joshi}
\author[a]{A. N. Ramprakash}
\author[a]{Chaitanya V. Rajarshi}
\author[a]{Rani S. Bhandare}

\affil[a]{Inter-University Centre for Astronomy and Astrophysics, Pune, India 411007}

\authorinfo{Further author information: (send correspondence to Sorabh Chhabra)\\Sorabh Chhabra: E-mail: sorabh.chhabra@gmail.com}

\begin{document} 
\maketitle

\begin{abstract}
The Speckle Imager via MUlti Layer Atmospheric Turbulence Object Reconstructor (SIMULATOR) is a lab-based testbed instrument developed to test for speckle correlation-based techniques in the optical regime. However, this instrument can be used as a testbed against post-processing techniques or algorithms like lucky imaging, phase diversity method etc. The SIMULATOR can emulate 3D atmospheric turbulence behaviour using a three-layer turbulence screen, giving the user command over important site characteristics like wind profile, global fried parameter, global isoplanatic patch, mid-layer and high-layer height effects etc. This testbed is unique in that it can mimic a broad range of site and telescope characteristics accurately without the need for manual intervention or tuning of parameters. The current version can handle a Field of View (FoV) of up to $0.3^{\circ}$, bandwidth ranges from 4860 to 6560 nm and can cover atmospheric turbulence heights up to 83 km.
\end{abstract}


\section{INTRODUCTION}
\label{sec:intro}  
\textbf{SIMULATOR} stands for \textbf{S}peckle \textbf{I}mager via \textbf{MU}lti \textbf{L}ayer \textbf{A}tmospheric \textbf{T}urbulence \textbf{O}bject \textbf{R}econstructor. It is a speckle imager that can mimic the exact characteristics of various observation sites as requested. This instrument has been made with the sole purpose of applying post-processing techniques like speckle correlation-based imaging\cite{beavers1989speckle,horch2004speckle} in image reconstruction under diverse conditions, including telescope cites turbulence strength, wind profiles, sky background variations etc. However, this instrument can be used as a testbed against post-processing techniques or algorithms like lucky imaging\cite{staley2014lucky,law2006lucky}, phase diversity method\cite{gonsalves2018phase} etc. The idea of this instrument was inspired by the MAPS (Multi-Atmospheric Phase screens and Stars) instrument, developed under the European Space Organisation (ESO) Very Large Telescope (VLT) team, solely to test the Multi-Conjugate Adaptive Optics (MCAO) project\cite{kolb2004maps}. With this instrument in hand, we can overcome the time request for a telescope which is extremely hard to get these days. We can even strengthen our methods/algorithms under investigation against various instrument errors. This project aims to mimic the three-dimensional evolving turbulence effect, thus stimulating turbulence behaviour of the atmosphere, covering all layers of the atmosphere up to 83 km.

\begin{figure}[h!]
    \centerline{\includegraphics[angle=0,width=0.99\textwidth]{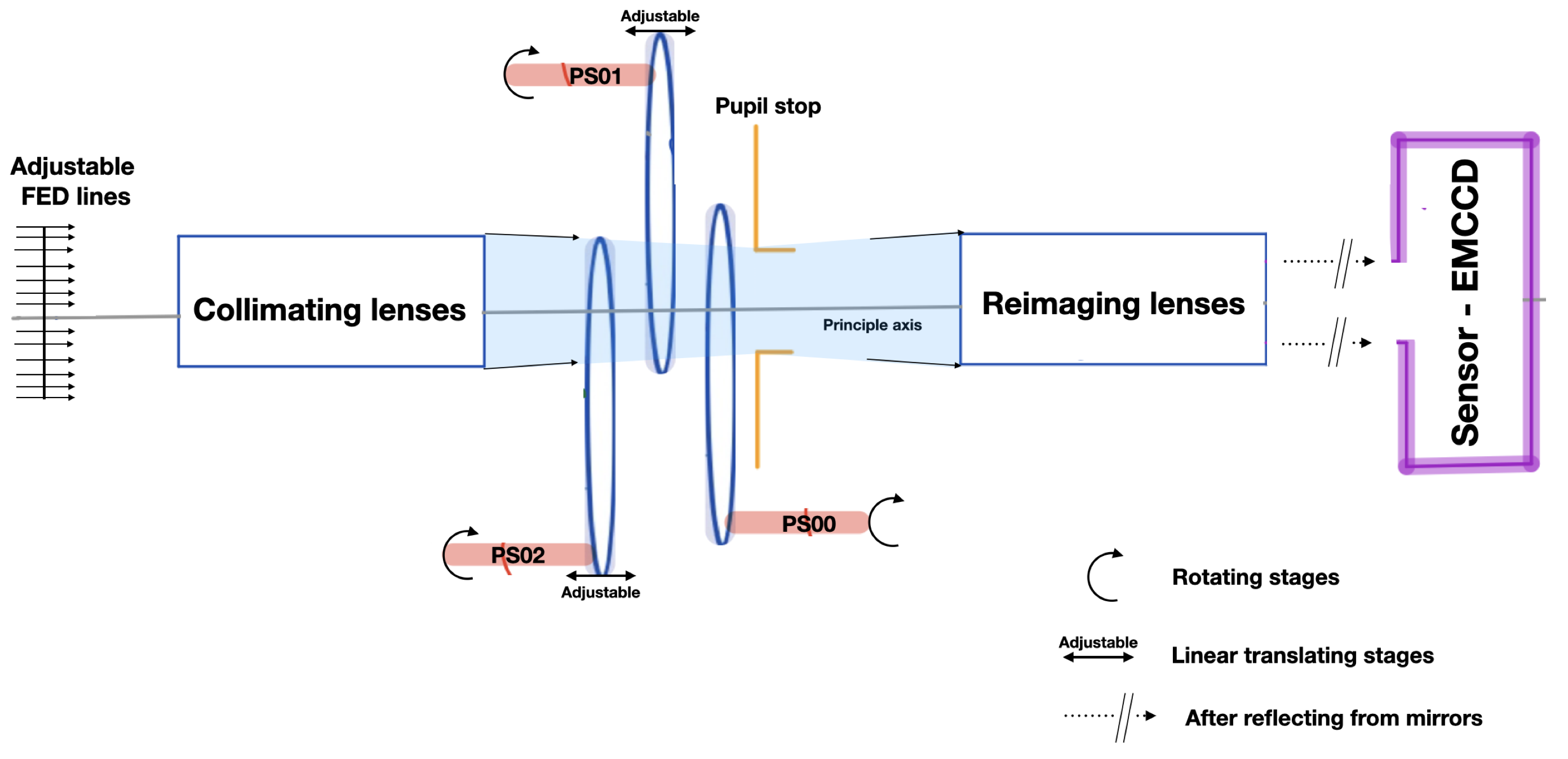}}
    \caption{Early stage conceptual design of SIMULATOR\label{fig:conceptual_design} }
\end{figure}

The early-stage conceptual design of SIMULATOR can be found in fig~\ref{fig:conceptual_design}. Upon user request, the source plane can provide adjustable fed lines from where the light originates. They can act as natural guide stars (NGS) depending on the field of view (FoV). A set of lenses combined to collimate the entire FoV. This collimated beam then passes through three turbulence phase screens before reaching the pupil plane. After exiting from the pupil stop, a combination of lenses re-images the incoming beam to the focus or sensor plane. With the help of mirrors, a slow converging beam can be brought to the camera plane. Here, the EMCCD camera is a sensor for fast frame capturing, rotating, and translating stages to simulate atmospheric turbulence effects by changing the atmosphere's $C_n^{2}$ profile.

\section{Characterization of SIMULATOR}
\label{sec:characterization}

\subsection{Turbulence phase screen}
\label{sec:SIMULATOR_phase_screen}

The turbulence in the atmosphere can be approximated by engraving the phase map profile on physical material disc named as turbulent phase screen, shown in fig.~\ref{fig:phase_screen}. The optical path difference gets generated for beams passing through different points across the field of observation. To design this model, we have adopted a three turbulence layer model. It means the three layers of the atmosphere can characterise an entire 3D turbulence volume. In this model, the turbulence strength profile parameter $C_n^2$ is considered constant across the plane for different height profiles\cite{wyngaard1971behavior}. Working with a single layer lowers this model's ability to mimic speckle features, including anisoplanatism and small-scale fluctuations effect at the pupil plane. These two effects can only be emulated if separate turbulence layers exist at both ends, ground and high altitude\cite{bos2012technique}. \\

\subsection{Scaling factors }
\label{sec:SIMULATOR_scalings}
\subsubsection{Site features}
\label{chap4:sec:SIMULATOR_scaling_telescope}

To incorporate an infinite distant long real imaging system within a finite-size optical table space of $0.8\times0.5$ m (based on initial calculations), we need to regulate all the parameters and scale them to smaller equivalent numbers. That's includes bringing telescope size $D$ to scaled-down version $d$, global fried parameter $r_0^{global}$ to $r_0^{scaled}$, turbulence wind velocity from $V(H)$ at height $H$ above telescope to scaled velocity $v(h)$ at scaled height $h$ above the pupil plane.\\

\subsubsection{Science features}
\label{chap4:sec:SIMULATOR_scaling_science}
 To emulate the features of a large telescope observatory within the lab, we need to do matching of the characteristics parameters in a small space within the optical bench. A speckle image can be characterised by simple parameters ratio number $D/r_0$, where $D$ is the size of the telescope and $r_0^{global}$ is the global fried parameter of any particular site. This ratio estimates the number of coherence patches through the telescope imprinted within a single speckle image. This number automatically becomes significant for a large observatory. For example, the SOAR Observatory in Chile corresponds to 27 at $0.5\mu m$ wavelength. Thus, any site characteristic features can be emulated through this instrument, given a provision is in place to play around with the size of the simulated telescope number and fried parameter. 
 
 This instrument aims to have the largest possible $D/r_0$ value. Thus, after giving thorough research, we decided to mimic telescope of size $D = 25 $ m in diameter with fried parameter of $r_0^{global} = 14.4 $ cm (average number for Paranal observatory site\cite{kolb2004maps}). This results in coherence patches count to $\approx 174$. Our research includes the maximum size of the turbulent phase screen that can be manufactured. Thus limiting maximum height up to which atmosphere can be covered within the imaging system, i.e maximum footprint size(shown in fig. \ref{fig:phase_screen}).

\subsubsection{Atmospheric height profile}
\label{chap4:sec:SIMULATOR_scaling_turbulence}
Our atmospheric turbulence ranges up to 100 km (roughly ) as measured from the telescope pupil plane. Thus, the ground layer location (PS00) plays a vital role in incorporating maximum turbulent height within the minor separation. Ideally speaking ground layer of the atmosphere should be situated next to the pupil plane. Physically this is impossible to achieve. For reference, the ground layer can be called to be located at 250 m above the telescope plane. The minimum separation feasible between PS00 and pupil stop (PS) is 0.5 mm. This constraint comes from the mechanical regulation of the model. The separation between the PS00 exit face and the exit face of PS must be 0.5 mm. A more appropriate solution has been adopted in the mechanical design and covered in section~\ref{sec:opto_final_design}. A simple derivation tells height of the atmosphere falls as square of the height above the telescope. Height $H$ and scaled height $h$ above the telescope pupil goes as follows\\

\begin{equation}
\frac{h}{H} = \Big(\frac{d}{D}\Big)^2  
\end{equation}

\begin{figure}[p]
\centering
   \begin{subfigure}{0.5\linewidth}
   \centering
   \includegraphics[width=\linewidth]{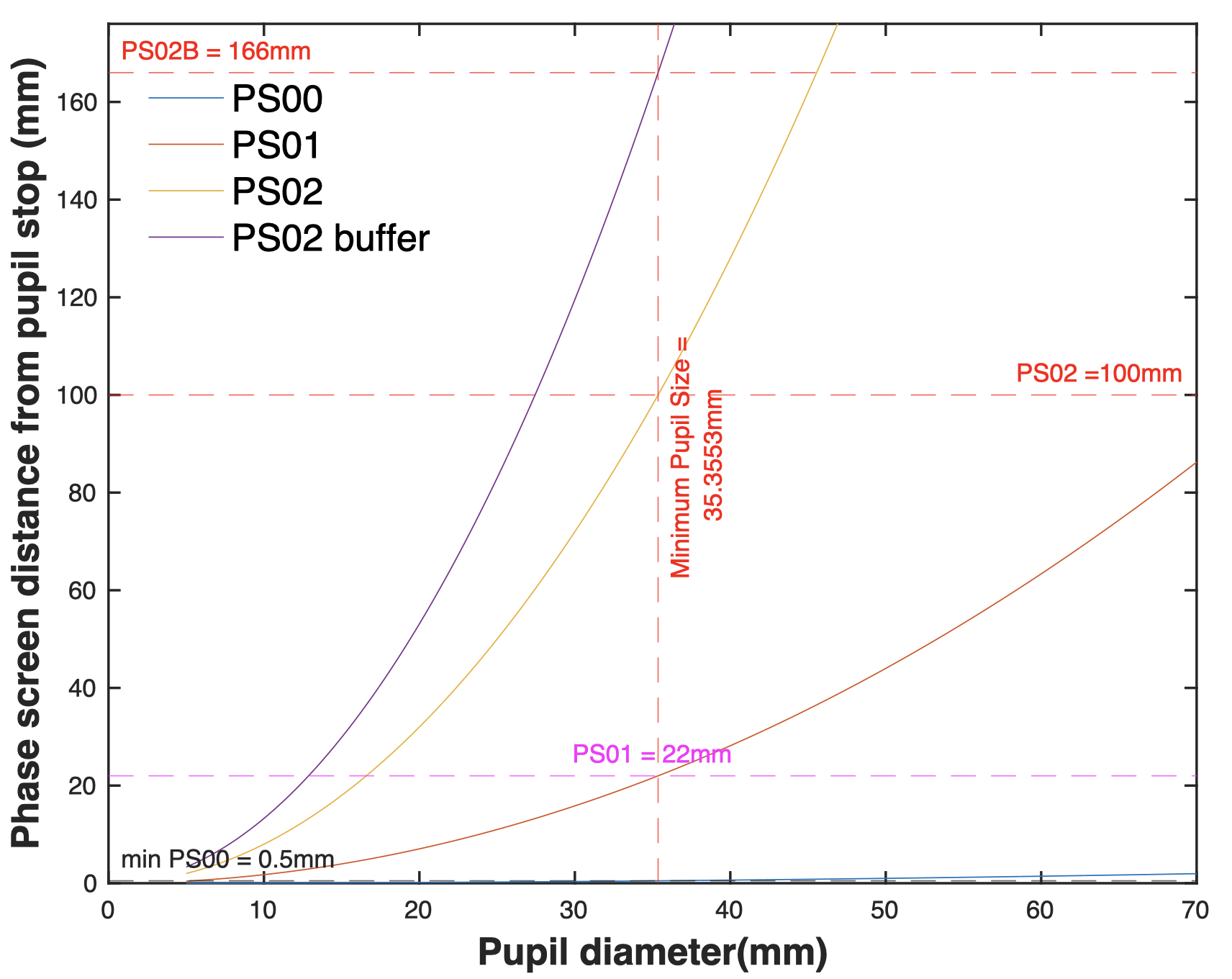}
   \caption{\label{fig:scaling_location}}
\end{subfigure}
\hfill
\begin{subfigure}{0.5\linewidth}
   \centering
   \includegraphics[width=\linewidth]{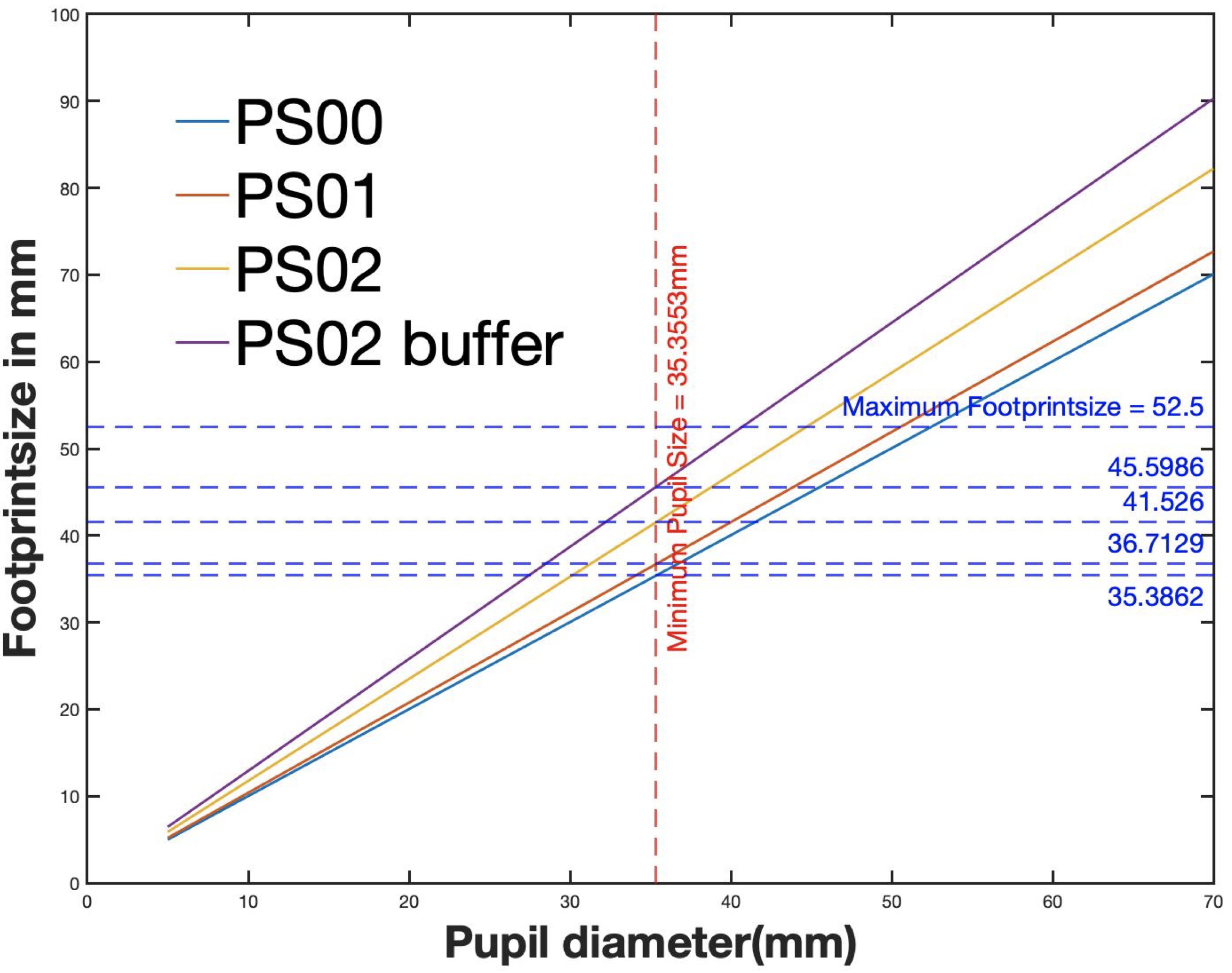}
   \caption{\label{fig:scaling_footprint} }
\end{subfigure}
\\[\baselineskip]
\begin{subfigure}[H]{0.5\linewidth}
   \centering
   \includegraphics[width=\linewidth]{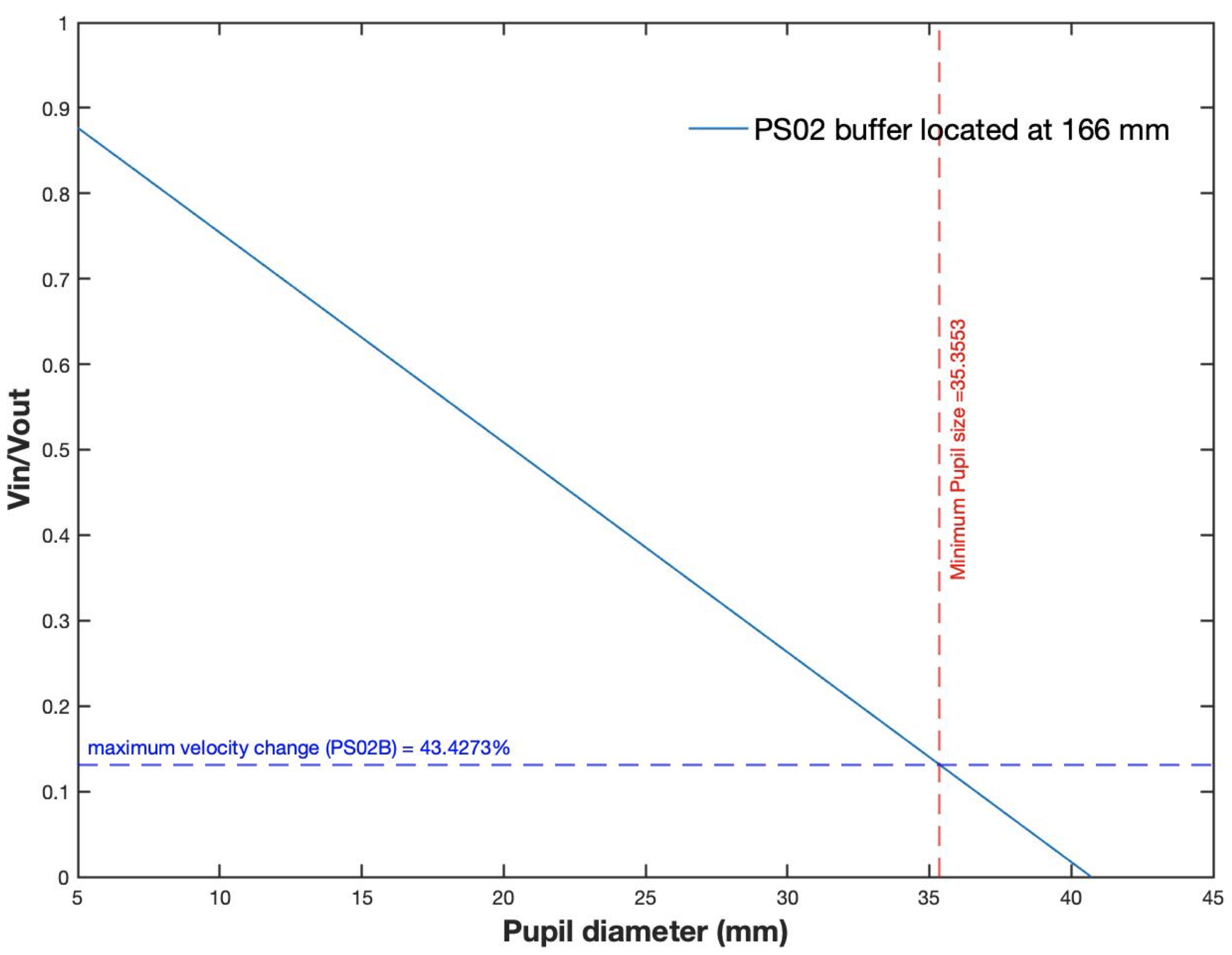}
   \caption{\label{fig:scaling_wind} }
\end{subfigure}
\centering
\caption{(a) Phase screen location for $d_{min} = 0.5$ mm , $D = 25 $ m. (b) Phase screen location based on FoV = $0.3^{\circ}$ , maximum footprint size at PS02 buffer = 52.5 mm. (c) Differential wind profile across PS02B versus pupil diameter. }
\end{figure}

Thus, for fixed aperture of size $D$ = 25 m, and minimum step size of h = 0.5 mm (corresponding H = 0.25 km), gives scaled pupil stop $d$ = 35.355 mm. Fig~\ref{fig:scaling_location}, plots pupil stop size versus phase screen distances from pupil stop. For min PS00 = 0.5 mm , and corresponding d = 35.355 mm, gives PS01 = 22 mm, PS02 = 100 mm and PS02B (buffer space for linear translation of PS02) = 166 mm. Here PS01, PS02 are mid and high altitude turbulence phase screen and  more details are covered in section~\ref{sec:SIMULATOR_phasescreen}.

\begin{figure}
    \centering
    \begin{subfigure}[b]{0.5\linewidth}        
        \centering
        \includegraphics[width=\linewidth]{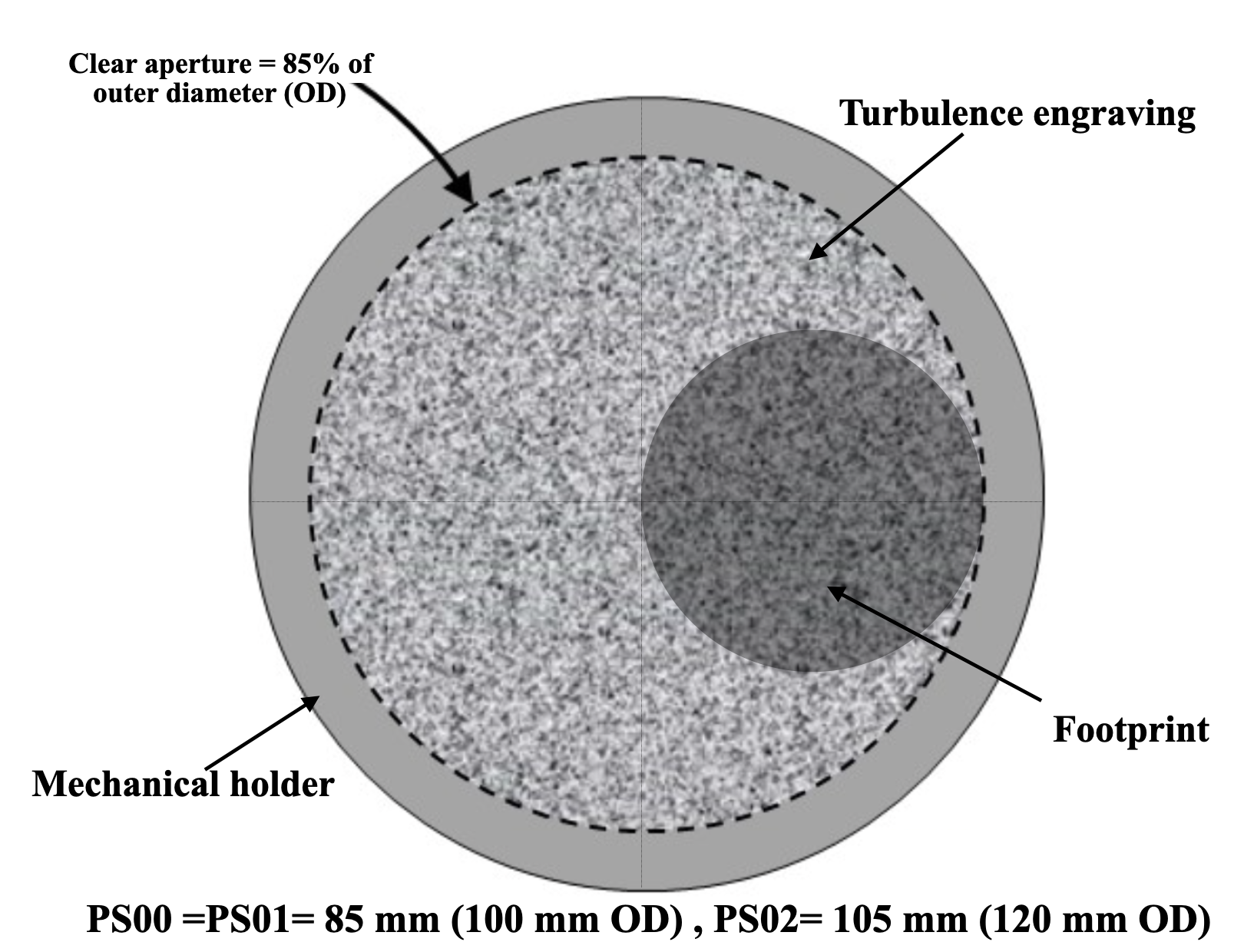}
        \caption{\label{fig:phase_screen}}
    \end{subfigure}
    \begin{subfigure}[b]{0.38\linewidth}        
        \centering
        \includegraphics[width=\linewidth]{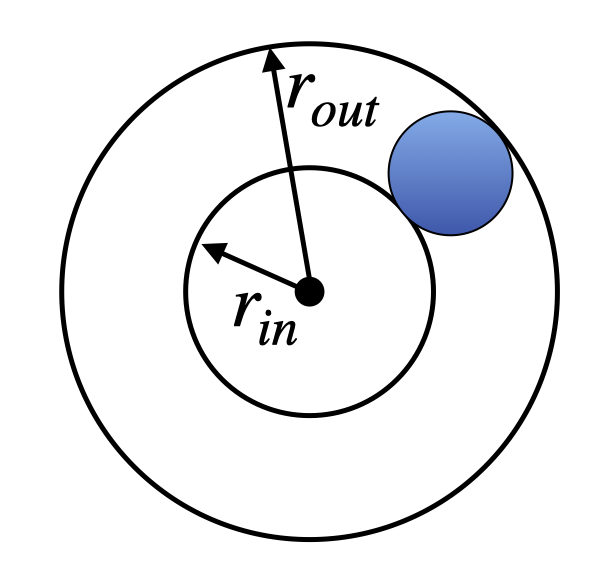}
        \caption{\label{fig:footprint}}
    \end{subfigure}
    \caption{(a) Turbulence phase screens PS00, PS01 and PS02 physical features. (b) Footprint impression on phase screen, where $r_{in}$ and $r_{out}$ are inner and outer radius. For our case, $r_{in}$ = 0}
    \label{fig:roc_curve}
\end{figure}

The three turbulent phase screens used for this experiment are of sizes PS00 = PS01 = 100 mm (85 mm clear aperture) and PS02 = 125 mm, each 1 mm thick are shown in fig~\ref{fig:phase_screen}. These phase screens are procured from Silios technologies, located in France. Thus maximum footprint allowed over each phase screens are PS00 = PS01$<$ 42.5 mm and PS02 $<$52.5 mm. Fig.~\ref{fig:scaling_footprint} shows plots for footprint size versus various pupil stop sizes. Thus, for PS00 = 35.38 mm ($<$ 45 mm), PS01 = 36.71 mm ($<$ 45 mm), PS02 = 41.53 mm ($<$ 52.5 mm)and PS02B = 45.59 mm ($<$ 52.5 mm). Calculation for footprint size $y$ goes as follows

\begin{equation}
\label{eq:footprint}
y = d+2h tan(\theta)
\end{equation}

where, FoV = 2$\theta$ scaled as 

\begin{equation}
\label{eq:fov}
\frac{\phi}{\theta} = \frac{d}{D}  
\end{equation}

where, $\phi$ is the scaled version of half of FoV. Thus, for 2$\theta = 0.3^{'}$ (Table.~\ref{tab:instrument_table}), $\phi = 1.76^{\circ}$ .

The wind gradient profile is the next important factor to incorporate within this instrument. The speckle imaging technique works within the frozen flow hypothesis, which means the speckle should be fully developed during an exposure period of a single speckle. Statistically, the atmosphere effects remain more or less the same except for the wind effect. For instance, when we model atmosphere, we assume a constant wind profile at a given height $h$ above the ground, just like the $C_n^2$ profile. It is impossible to incorporate this into our system because extended phase screens (ranging from 100 mm to 125 mm in OD) will experience a differential wind profile across the area of interest (or footprint). However, this can be minimised by bringing phase screens closer to each other, thus helping to downsize footprint size. This is not fruitful given our desired goals due to the limited coverage of turbulence layers of the atmosphere. We aim to keep differential speed less than 50\% across the footprint region. Data also shows for VLT 8 m telescope aperture average number is close to 50\%\cite{kolb2004maps}. For three-phase screens, we have picked wind velocity PS00 = 7 $\pm$ 3 $ms^{-1}$, PS01 = 30 $\pm$ 15 $ms^{-1}$ and PS02 = 60$\pm$ 30 $ms^{-1}$\cite{kolb2004maps}. The scaling factor for wind velocities goes linearly with pupil size as,

\begin{equation}
\label{eq:wind_scaling}
\frac{V}{v}= \frac{D}{d}  
\end{equation}

Thus, velocity by which phase screen needs to be constantly rotated are PS00 = $9.9\pm4.2 mm/s (7\pm 3 m/s) 
$, PS01 = $42.4\pm21.2 mm/s (30\pm 15 m/s)$ 
 and PS02 = $84.9\pm42.4 mm/s (60\pm 30 m/s)$. The following equations have been employed to find the gradient change across the PS02 screen. 

\begin{equation}
\label{eq:wind_1}
\frac{V_{in}}{V_{out}}= \frac{r_{in}}{r_{out}}  
\end{equation}

\begin{equation}
\label{eq:wind_2}
\frac{V_{in}}{V_{out}} = \frac{r_{out}-y}{r_{out}}
\end{equation}

where footprint $y$ is used from eq.~\ref{eq:footprint} and also shown in fig~\ref{fig:footprint}. Fig~\ref{fig:scaling_wind} plots differential velocity plot ($V_{in}/V_{out}$) versus pupil stop diameter. For pupil stop size of 35.355 mm, differential wind across footprint PS02B is within the tolerance level of the atmosphere ($<$ 50\%). Results are shown only for the case of PS02B, which can encapsulate maximum error because it is located farthest from the pupil stop.                

\subsection{Fried parameter to PS00, PS01 and PS02}
\label{sec:SIMULATOR_fried}
As per paranal site characteristics, the global fried parameter $r_0^{G}$ is 14.4 cm at 0.5$\mu m$ in median seeing condition \cite{kolb2004maps}. Thus, a scaled version of the fried number goes as

\begin{equation}
\label{eq:fried_scaling}
\frac{r_0^{G}}{r_{0}^{g}} = \frac{D}{d}
\end{equation}

Thus producing scaled global $r_{0}^{g}$ value of 0.203 mm. Next, we aim to parametrise each phase screen fried number for given global $r_{0}^{g}$. Table~\ref{tab:phase_screen_fried_parameter} has been used to find fried parameters for each independent phase screens. The $i^{th}$ layer fried parameter in terms of $C_n^{2}$ profile is given as\cite{roggemann1996imaging}  

\begin{equation}
\label{eq:ith_layer}
r_{0_i} = [0.423\kappa^2C_{n_i}^2\Delta h_i]^{-3/5}
\end{equation}

where, $\Delta h_i$ marks distance between $i^{th}$ and $(i-1)^{th}$ layer. We can write the desired optical field parameters isoplanatic angle $\theta_{0}$ and log amplitude variance $\sigma
_{\chi}^2$, in terms of the phase-screen $r_{0_i}$ values as\cite{roggemann1996imaging}

\begin{equation}
\label{eq:ith_layer_fried}
r_0 = \Bigg[\sum_{i=0}^{i=n}r_{0_i}^{-5/3}(h/\Delta h)^{5/3}\Bigg]^{-3/5}
\end{equation}

\begin{equation}
\label{eq:ith_layer_iso}
\theta_0 = \Bigg[6.8794L^{5/3}\sum_{i=0}^{i=n}r_{0_i}^{-5/3}(1-h_i/\Delta h)^{5/3}\Delta h_i\Bigg]^{-3/5}
\end{equation}

\begin{equation}
\label{eq:ith_layer_sigma}
\sigma_{\chi}^2 = 1.331\kappa^{-5/6}\Delta h^{5/6}\sum_{i=0}^{i=n}r_{0_i}^{-5/3}(h_i/\Delta h)^{5/6}(1-h_i/\Delta h)^{5/6}\Delta h_i
\end{equation}

Next step is to find the individual fried parameters $r_0^{PS00}$, $r_0^{PS01}$ and $r_0^{PS02}$. Above equations eq.~\ref{eq:ith_layer_fried} and eq.~\ref{eq:ith_layer_sigma} can be rewritten in terms of matrix notation, and can be solved using fmincon function within MATLAB on eq.~\ref{eq:fmincon} by imposing constraint on log amplitude variance cannot exceed more than 20\% by each screen\cite{schmidt2010numerical} and remaining fixed parameters from table.~\ref{tab:instrument_table}. 

\begin{equation}
\label{eq:fmincon}
\resizebox{0.95\hsize}{!}{$
\left[\begin{array}{c}
\left(\hat{r}_{0}\right)^{-5 / 3} \\
\frac{\hat{\sigma}_{x}^{2}}{1.331 \kappa^{-5 / 6} L^{5 / 6}} 
\end{array}\right]=\left[\begin{array}{cccc}
\left(h_{1} / L\right)^{5 / 3} & \left(h_{2} / L\right)^{5 / 3} & \ldots & \left(h_{N} / L\right)^{5 / 3} \\
\left(h_{1} / L\right)^{5 / 6}\left(1-\frac{h_{1}}{L}\right)^{5 / 6} & \left(h_{2} / L\right)^{5 / 6}\left(1-\frac{h_{2}}{L}\right)^{5 / 6} & \ldots & \left(h_{N} / L\right)^{5 / 6}\left(1-\frac{h_{N}}{L}\right)^{5 / 6} 
\end{array}\right] \times\left[\begin{array}{c}
r_{01}^{-5 / 3} \\
r_{0_{2}}^{-5 / 3} \\
\ldots r_{0_{N}}^{-5 / 3}
\end{array}\right]$}
\end{equation}

$N$ represents a number of phase screens (3 for our case), $\kappa$ is the wavenumber. The output from \textit{fmincon} are $r_0^{PS00}$ = 0.308 mm, $r_0^{PS01}$ = 0.3039 mm and $r_0^{PS02}$ = 0.3366 mm. The final output can be cross-checked using eq.~\ref{eq:ith_layer_iso} and verified against theoretical isoplanatic results. Our calculation shows maximum of 0.04\% error remains in isoplanatic measurements.

\section{PRELIMINARY DESIGN}
\label{sec:concept}
\begin{figure}[h!]
    \centerline{\includegraphics[width=0.6\textwidth]{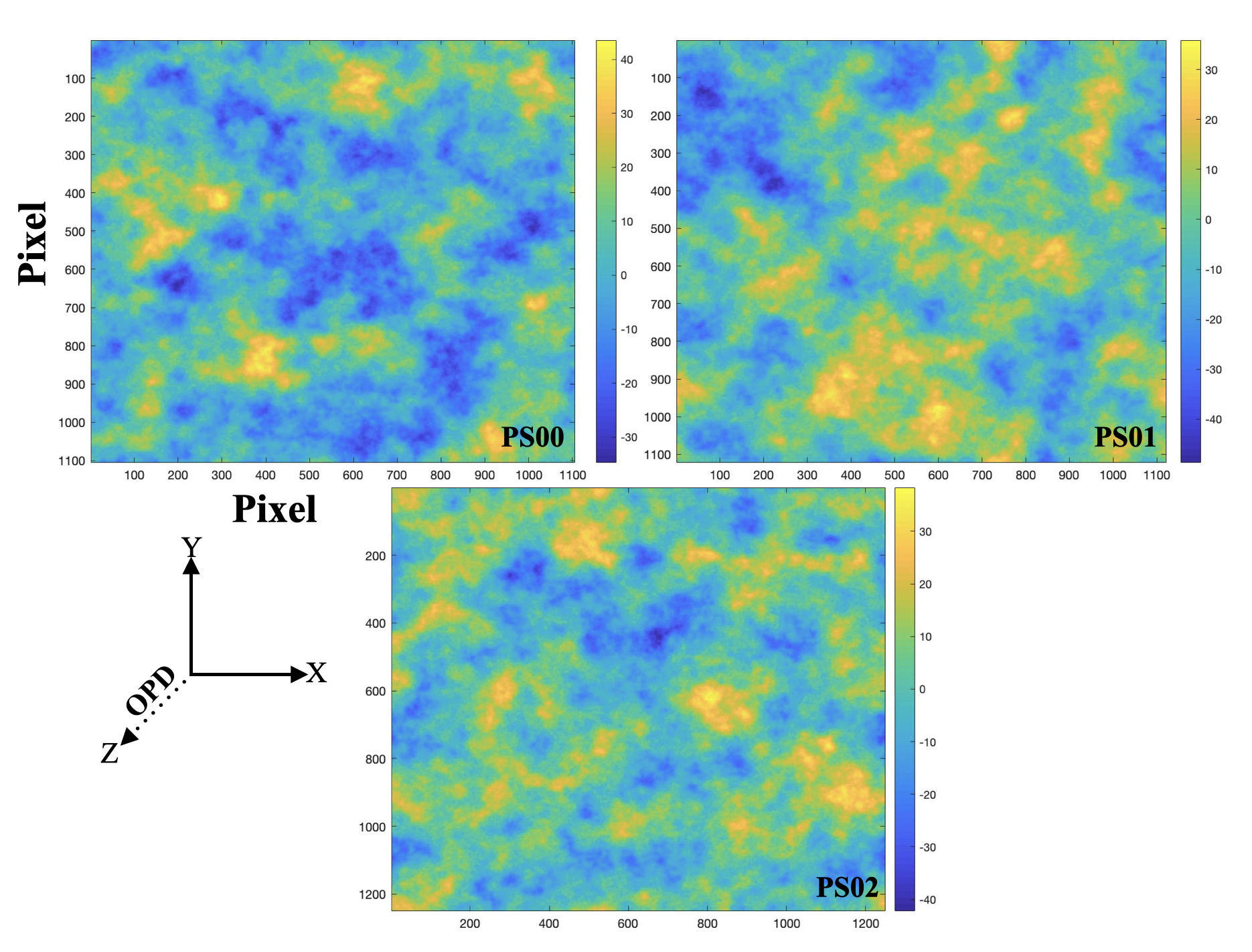}}
    \caption{Phase screen maps of PS00, PS01 and PS02 respectively\label{fig:phase_maps}  }
\end{figure}

Our design comprises three stages. The first stage demands three sets of phase screens engraved with turbulence properties, as per fig.~\ref{fig:phase_maps}. The second stage covers a set of optics required to produce a collimated beam from natural guide star/point sources located at the source plane for the desired FoV. The third stage requires a set of optics for focusing distorted wavefront, slow converging beam towards camera plane for diffraction-limited results with minimal seidal coefficients (in fig.~\ref{fig:seidel}).

Table~\ref{tab:instrument_table} contains a list of parameters that have been incorporated within the instrument during its design. We must test our speckle imaging technique against a broad wavelength range of 150 nm in the optical regime. The requisite space of 20 mm at the source plane corresponds to the $0.3^{'}$ of FoV (or $3.534^\circ$ scaled FoV ). Thus, the input f-number of 4.95 is chosen for that reason. Since this system has been designed to have a diffraction-limited result for a telescope size of 25 m, it thus requires a slow converging beam for output. Output F-number of 51 corresponds to a back focal length of 1800 mm, and this has been achieved using multiple fold mirrors for compactness. The following subsections will explain the zemax output design for various subsystems within the instrument. \\   

\subsection{Fixing phase screens}
\label{sec:SIMULATOR_phasescreen}

Fig~\ref{fig:phase_screen_layout} contains first stage information, which contains detailed information of three-phase screen location (shown in table.~\ref{fig:phase_screen_layout} \& \ref{tab:phase_screen_fried_parameter}). Each phase screen is placed off-centred w.r.t. principle axis of the beam because of the symmetric nature of turbulence profile engraved to it, as shown in fig~\ref{fig:phase_screen} and. We want the beam to be passed only through one-half of the turbulent phase. Corning C7980 is the material the manufacturer uses for making, and each phase screen is only 1 mm thick. PS00 and PS01 are 100 mm in diameter and PS02 is 125 mm in diameter. Thus, these numbers are initially put within the zemax at the very early design stage and kept fixed. Although for design purposes, phase screens PS01 and PS02 have been kept fixed to 22 mm and 100 mm from pupil stop. A provision in mechanical design is made for the movement of the phase screen by $\pm$ 30 mm and $\pm$ 15 mm. This instrument must bring turbulence profile variability for testing purposes (details covered in table \ref{tab:phase_screen_variability}).

The memory cell \cite{roddier1981atmospheric,roddier1982isoplanatic} or fried parameter is another critical parameter that gets affected by variability in phase screen location. Memory cell size is about 1.5 times larger than the fried parameter. For configuration where PS01 and PS02 have located at 22 km and 83 km, respectively, memory cell size turns out to be $0.0702^\circ$ ( entire FoV: $3.534^\circ$ ). Thus, leaving a total number of memory cells across one axis around 50 or 2500 cells within the $10\times 10$ mm region of the object plane. Each cell turns out to be 400 $\mu$m in size at the source plane. Atleast a couple of source is required to be fitted within 400 $\mu$m for testing.

\subsection{Collimating optics}
\label{sec:SIMULATOR_s2p}
After fixing up phase screens at desired locations, the next task is to focus light from $10\times 10$ mm within the source plane that must be collimated before entering the phase screen model. With the help of zemax, we have used a set of four custom-made collimating lenses (CL) for this particular task CL1, CL2, CL3 and CL4. For design optimisation, two important parameters had taken under consideration. Footprint shift per unit memory cell (or fried parameter) across three phase screens and pupil shift due to chromatic effects. Ideally, wavefront or light exiting from CL4 (collimated) originated from one ideal point source should overlap for the entire wavelength band. But, due to different refractive materials, each light ray bends or interacts with lenses differently (governed by Cauchy's equation\cite{roddier1999adaptive}). This leads to a shift in wavefront w.r.t. to the central propagation wavelength. This shift in wavefront w.r.t. to central wavelength is called footprint shift. Zemax helps in eliminating this shift as much as close to ideal figures. But due to the restrained number of lens catalogues, achieving an ideal figure is almost impossible. Thus a tolerance in final results is tested against this shift, and a compromise has been made. Our scientific goal requires that incoming light passes through the same coherence patch of sky or, in other words, speckles must be fully developed during exposure. As shown in table~\ref{tab:table_fixed_2}, we manage to bring footprint shift down to 5.7\% $r_0$ for PS02 i.e less than 19 $\mu$m, 0.02\% $r_0$ for PS01 i.e less than 0.06 $\mu$m and 0.1\% $r_0$ for PS00 i.e less than 0.34 $\mu$m. Fig~\ref{fig:pupil_footprint} shows footprint diagram covering entire FoV of $10\times 10$ mm and entire broadband range of 0.486 to 0.656 $\mu$m.

\begin{figure}[h!]
    \centerline{\includegraphics[angle=0,width=0.9\textwidth]{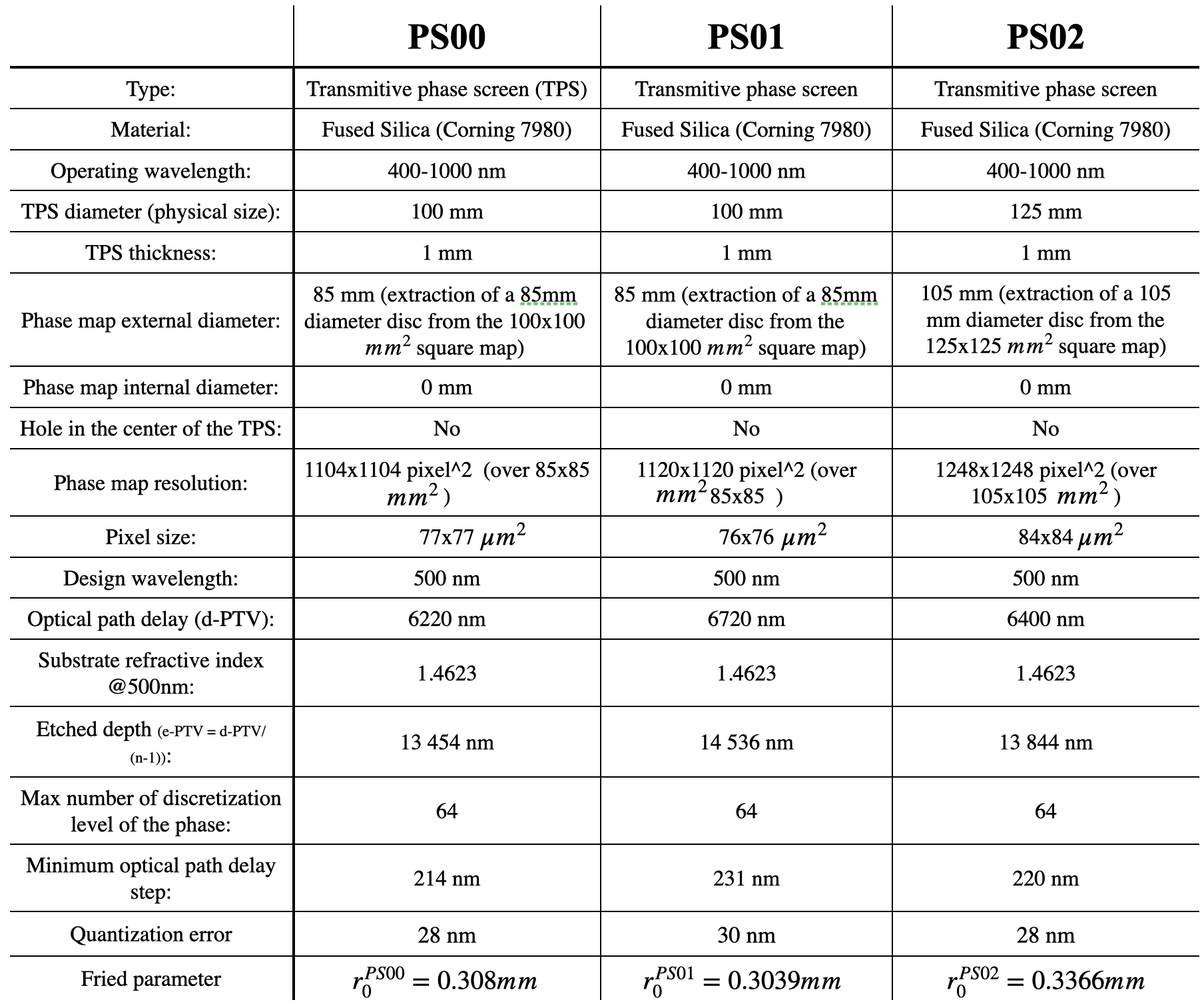}}
    \caption{Technical description of all three phase screens PS00, PS01 and PS02.\label{fig:phase_screen_layout} }
\end{figure}

\begin{figure}[h!]
    \centerline{\includegraphics[angle=0,width=0.8\textwidth]{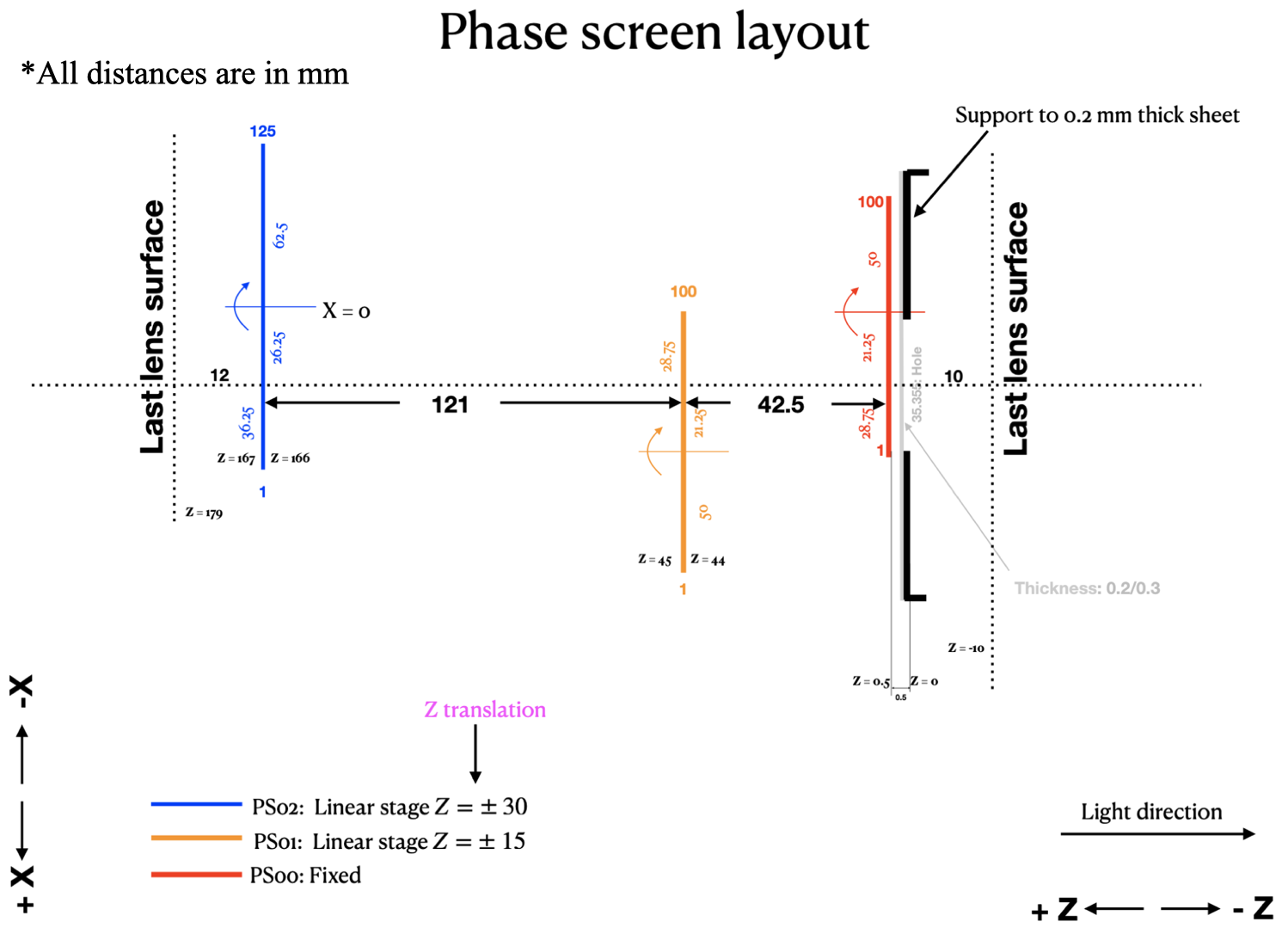}}
    \caption{Proposed phase screen layout for PS00, PS01 and PS02. All distances are in mm\label{fig:phase_screen_layout} }
\end{figure}

\subsection{Re-imaging optics}
\label{sec:SIMULATOR_p2c}
Our next scientific target is a diffraction-limited result at the focal or camera plane. Our sensor area is only $8.2\times 8.2$ mm ( EMCCD camera iXon Ultra 897, covered in tab ~\ref{tab:EMCCD_1} ). We finally decided to focus on smaller FoV $1\times 1$ mm  (corresponding plate scale of 1.9 arcmin/mm, output F-number 51 and back focal length of 1805 mm). With the help of zemax, the best combination of spherical lenses is selected by playing around with merit function parameters. Our major focusing parameters include WFNO (working f-number), EFLX - Effective focal length ( back focal length), and optimisation parameter covers RMS spot radius (centroid) at the focal plane. Finally, we ended up with four imaging lenses IL1, IL2, IL3 and IL4. The separation from the IL4 lens's last surface to the camera plane accounts for about 1333 mm. We have used five-fold mirrors (M1-M5) to make the system compact. However, depending on the optomechanical constraint (e.g. space constraint within an optical bench ), this design can be readjusted without adjusting other lenses combination (shown in fig.~\ref{fig:OM_mirror}).

\section{Final design}
\label{sec:final_design}

\begin{figure}[h!]
    \centerline{\includegraphics[angle=0,width=0.9\textwidth]{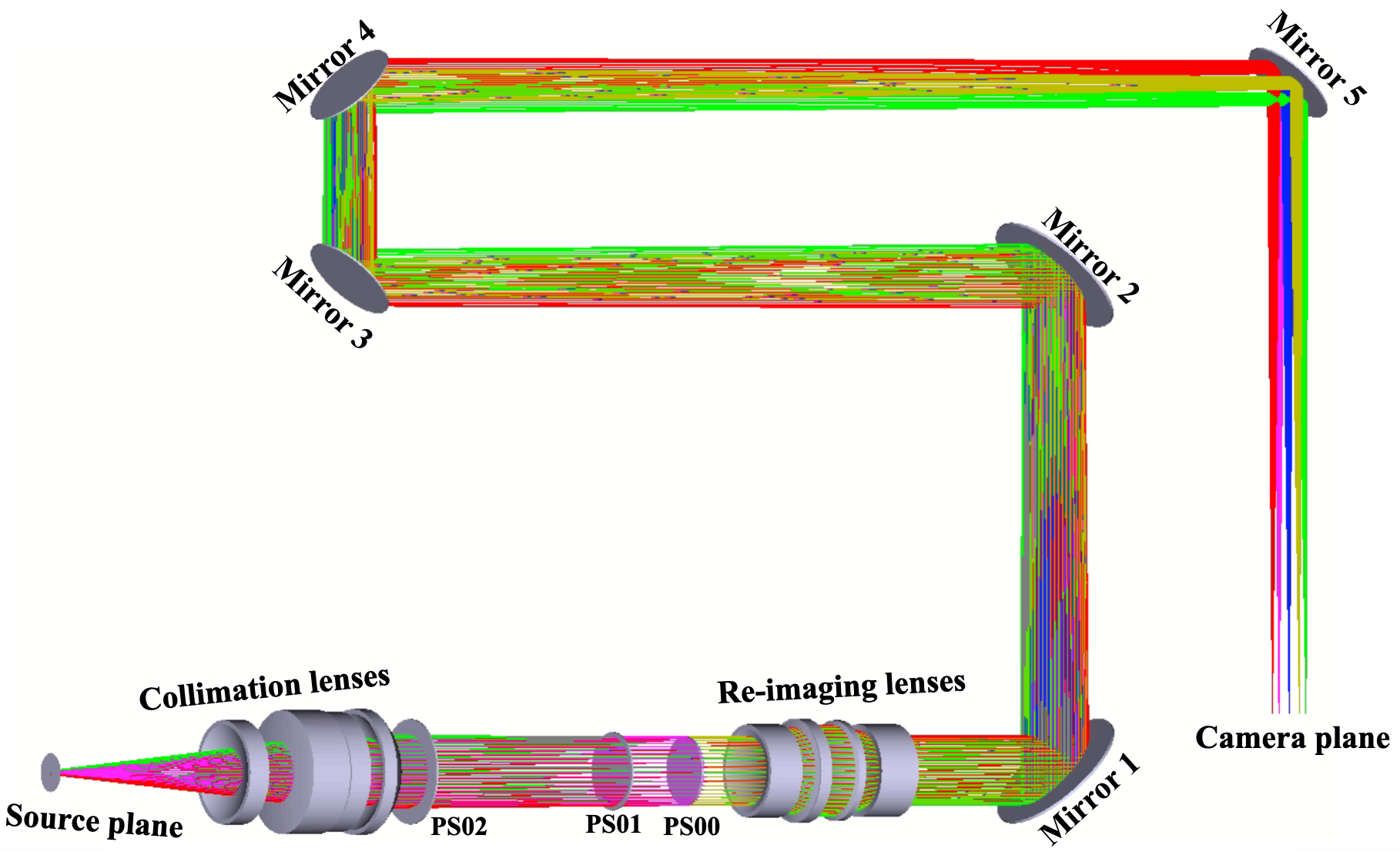}}
    \caption{Final zemax design for SIMULATOR\label{fig:final_design} }
\end{figure}

\begin{figure*}
\centering

\begin{subfigure}[t]{0.45\textwidth}
\centering
\includegraphics[width=\textwidth]{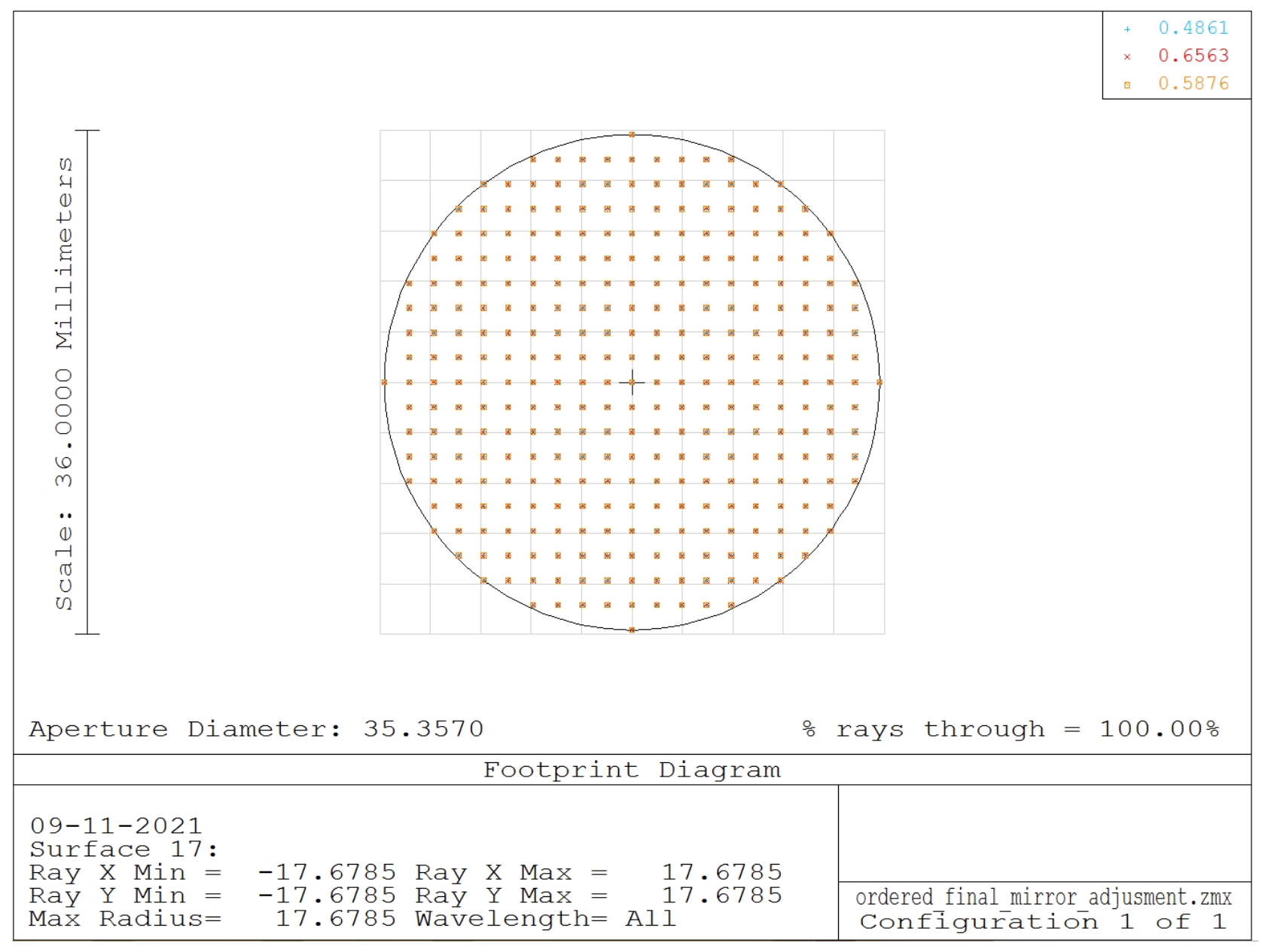}
\caption{\label{fig:pupil_footprint} Footprint diagram at pupil plane for entire field of view $10\times 10$ mm, covering entire wavelength range 0.486 to 0.656 $\mu$m}
\end{subfigure}%
\hfill
\begin{subfigure}[t]{0.45\textwidth}
\centering
\includegraphics[width=\textwidth]{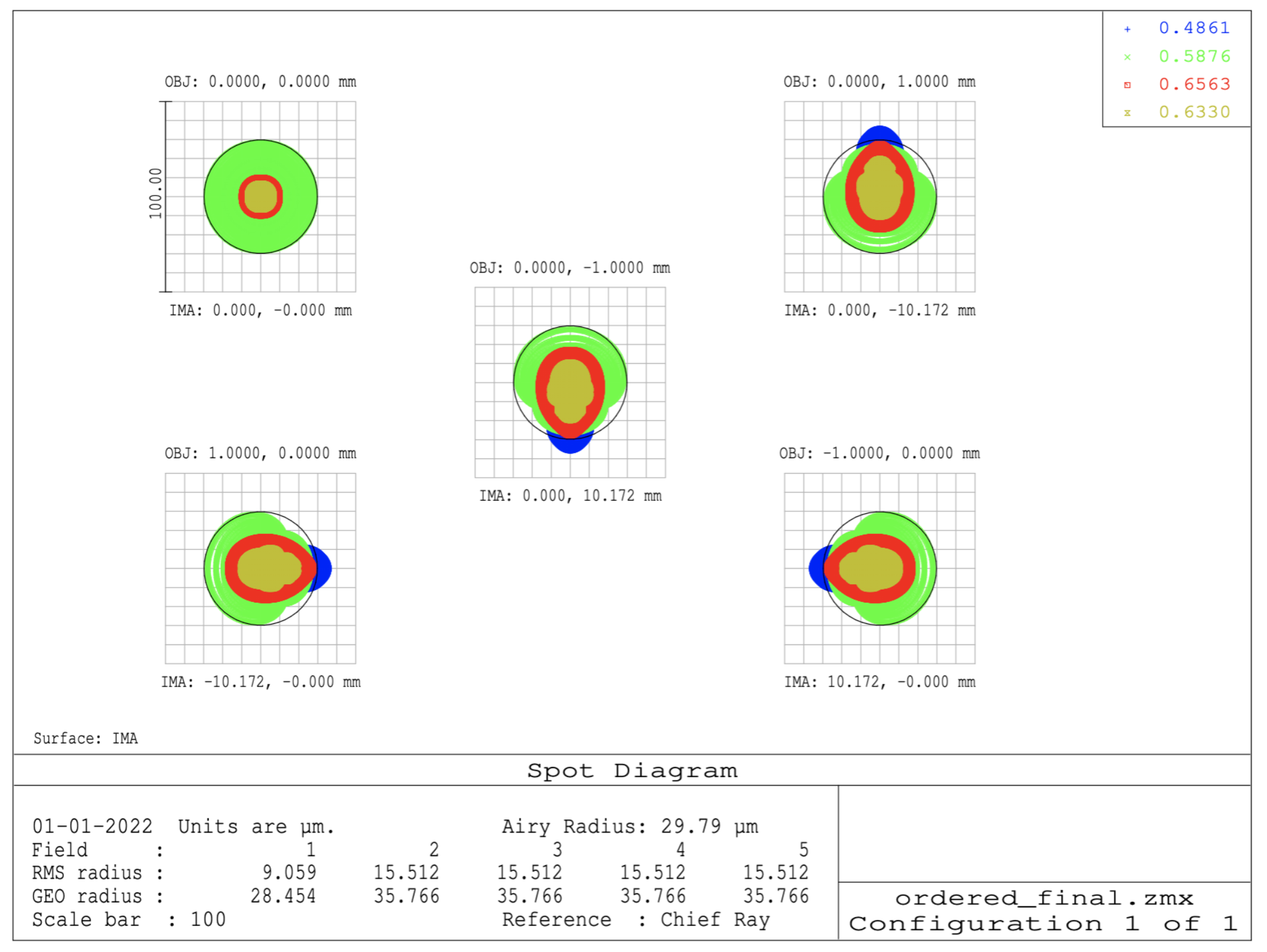}
\caption{\label{fig:spot_diagram} Spot diagram for 2 mm FoV, covering bandwidth range of 0.4861 to 0.6563 $\mu$m }
\end{subfigure}

\bigskip 

\begin{subfigure}[t]{0.45\textwidth}
\centering
\includegraphics[width=\textwidth]{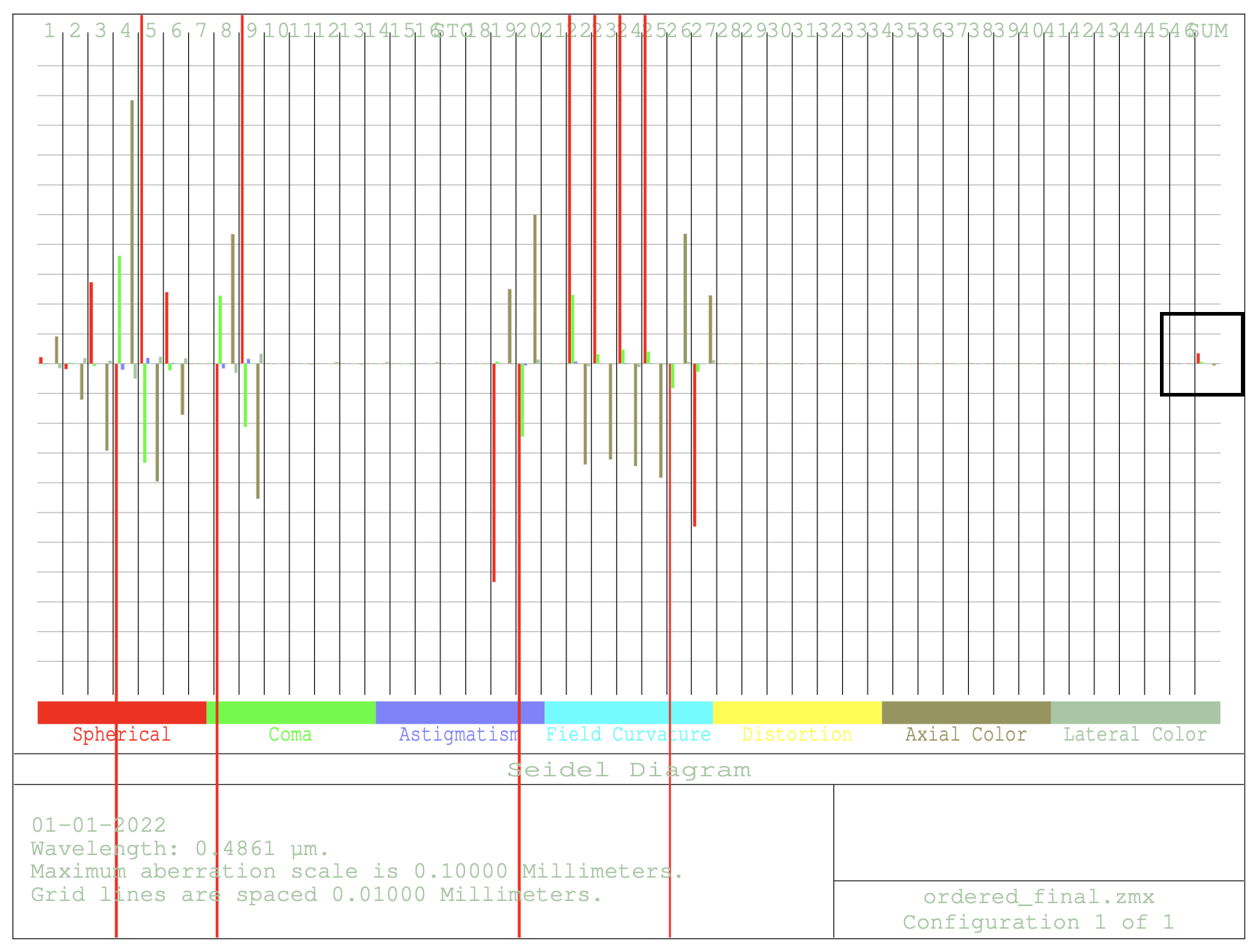}
\caption{\label{fig:seidel} Net seidel coefficients at camera plane, each square dimensions are 0.01$\times$0.01 mm}
\end{subfigure}%
\hfill
\begin{subfigure}[t]{0.45\textwidth}
\centering
\includegraphics[width=\textwidth]{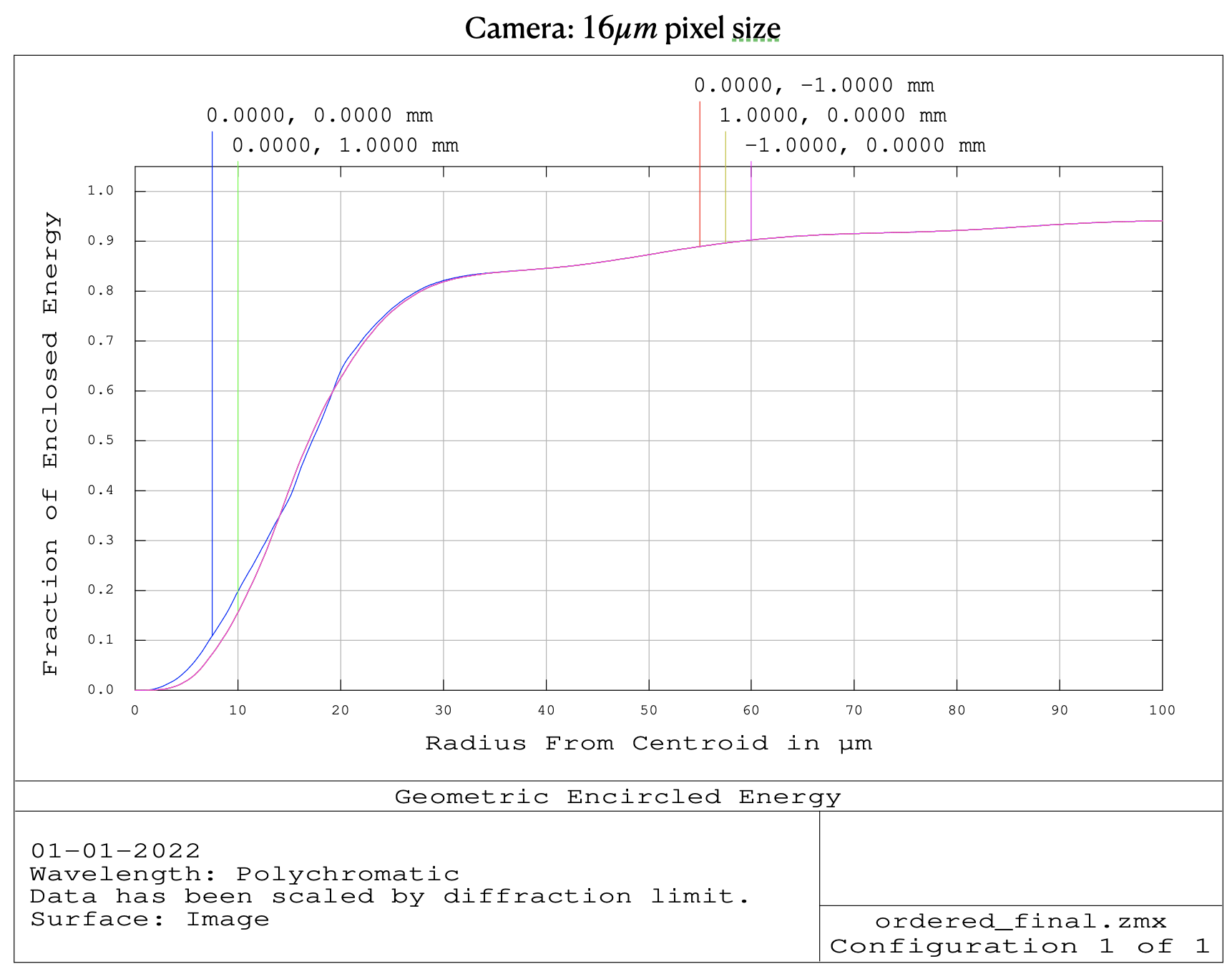}
\caption{\label{fig:encircled} Radius of centroid in $\mu$m vs fraction of geometric encircled energy for 2 mm FoV}
\end{subfigure}

\caption{Final design residual errors in terms of footprint shift, rms spot radius, net seidel coefficients and geometric encircled energy, within FoV 2mm }
\label{fig:final_design_errors}
\end{figure*}

Fig.~\ref{fig:final_design} shows the final zemax design of instrument SIMULATOR. Fig.~\ref{fig:final_design_errors} covers final residual errors in footprint shift, RMS spot radius, net seidel coefficients and geometric encircled energy within FoV 2mm. RMS spot radius for the outer most field is 15.5 $\mu m$, which is well within airy disc radius 29.79 $\mu m$ at 0.5 $\mu m$. Maximum aberration scale is 0.1 mm at 0.486 $\mu m$. 80\% of geometric encircled energy can be covered within 3.75$\times$3.75 pixels. 

\begin{table}[h]
    \begin{subtable}[h]{0.45\textwidth}
        \centering
        \resizebox{1\textwidth}{!}{%
        \begin{tabular}{l | l | l}
        \textbf{Item}                                                  & \textbf{Value}                                                         & \textbf{Comments}           \\
        \hline \hline
        { $r_{0}^{g}$}        & { 0.204 mm}                 & { scaled fried parameter}                                                                        \\ \hline
{ $\bar{\lambda}$}    & { 0.5$\mu$ m}               & { central wavelength}                                                                           \\ \hline
{D}                  & {25 m}                     & { aperture size}                                                                               \\ \hline
{ d}                  & { 35.355 mm}                & { pupil stop size}                                                                              \\ \hline
{height scaling}     & {2 $\mu$m}                 & { $\Big(\frac{d}{D}\Big)^2$}                                                                    \\ \hline
{h\_PS00}            & {0.5 mm}                   & { H\_PS00 = 0.25 km actual height}                                                              \\ \hline
{ h\_PS01}            & { 22 mm}                    & { H\_PS01 = 11 km, actual height}                                                               \\ \hline
{ h\_PS02}            & { 100 mm}                   & { H\_PS02 = 50 km, actual height}                                                               \\ \hline
{ L}                  & { 166 mm}                   & { \begin{tabular}[c]{@{}c@{}}Total length of atmosphere, \\ corresponds to 83 km.\end{tabular}} 
       \end{tabular}}
       \caption{}
       \label{tab:phase_screen_fried_parameter}
    \end{subtable}
    \hfill
    \begin{subtable}[h]{0.45\textwidth}
        \centering
        \resizebox{1\textwidth}{!}{%
        \begin{tabular}{l | l | l}
        \textbf{Item}                                                  & \textbf{Value}                                                         & \textbf{Comments}           \\
        \hline \hline
         
Wavelength range                                               & 0.486 - 0.656$ \mu$m                                                   & Optical range               \\ \hline
FOV:                                                           & 0.3 arcmin ($3.534^\circ$)                                             & Scaled  FoV in brackets     \\ \hline
\begin{tabular}[c]{@{}c@{}}F\#\\  \\ Input/Output\end{tabular} & \begin{tabular}[c]{@{}c@{}}Input : 4.95\\  \\ Output : 51\end{tabular} &                             \\ \hline
Plate scale                                                    & 1.9 arcmin/mm                                                          & $\lambda\times$ F\#(output) \\ \hline
Phase Screen Thickness                                         & 1 mm                                                                   & manufacturing constraints   \\ \hline
No. of screens                                                 & 3                                                                      & three layer turbulent model \\ \hline
Screen positions                                               & 0.25/11/50 km                                                          & Flexible upto 83 km         \\ \hline
Pupil Stop Size                                                & 35.355 mm                                                              & Fixed, as discussed earlier \\ \hline

        \end{tabular}}
        \caption{}
        \label{tab:instrument_table}
     \end{subtable}
     
     \caption{a) Parameters assigned to evaluate fried parameters for each phase screen, b) Important characterisations for instrument designing }
     
\end{table}

\begin{table}[]
\centering
\resizebox{1\textwidth}{!}{%
\begin{tabular}{|c|c|c|}
\hline
\textbf{Parameter}                                                                                                    & \textbf{Value}                                     & \textbf{Comment}                                                                             \\ \hline
\begin{tabular}[c]{@{}c@{}}2nd screen range  16 - 44 mm (8-22 km);\\  {[}$r_0^{g}$, $D/r_0^{g}${]}\end{tabular}       & {[} 0.201- 0.217 mm, 176 - 163 mm (8 - 22 km){]}   & moving only PS01 screen                                                                      \\ \hline
\begin{tabular}[c]{@{}c@{}}3nd screen range 44 - 166 mm (22 - 83 km);\\  {[} $r_0^{g}$,  $D/r_0^{g}$ {]}\end{tabular} & {[} 0.186 - 0.216 mm, 190 - 163 mm (22 - 83 km){]} & moving only PS02 screen                                                                      \\ \hline
$D/r_0^{g}$ over allowed range                                                                                        & 152 - 193                                          & \begin{tabular}[c]{@{}c@{}}193: 8/22 km (PS01/PS02)\\ 152: 22/83 km (PS01/PS02)\end{tabular} \\ \hline
Simulated telescope size                                                                                              & 21.8 - 27.79                                       &                                                                                              \\ \hline
\end{tabular}%
}
\caption{Variability in the simulated telescope size with changing phase screen positions}
\label{tab:phase_screen_variability}
\end{table}

\begin{table}[h]
    \begin{subtable}[h]{0.35\textwidth}
        \centering
        \resizebox{1\textwidth}{!}{%
        \begin{tabular}{l | l }
        \textbf{Parameter}                                                &\textbf{Value}                                                                  \\
        \hline \hline
        \begin{tabular}[c]{@{}l@{}}Chromatic effect :\\  Pupil Shift (total)\end{tabular}        & \begin{tabular}[c]{@{}l@{}}\textless 58$\mu$m  (28\% $r_0$) (across 20 cm FoV)\\  \textless 23 $\mu$m  (11\% $r_0$) (across 4 cm FoV)\end{tabular}                                        \\ \hline
\begin{tabular}[c]{@{}l@{}}Footprint Shift per cell:\\  In footprint planes\end{tabular} & \begin{tabular}[c]{@{}l@{}}PS02   : \textless 19 $\mu$m ( 5.7\% $r_0$ )\\  PS01   : \textless 0.06 $\mu$m ( 0.02\% $r_0$ )\\  PS00   : \textless 0.34 $\mu$m ( 0.1\% $r_0$ )\end{tabular} 
       \end{tabular}}
       \caption{}
       \label{tab:table_fixed_2}
    \end{subtable}
    \hfill
    \begin{subtable}[h]{0.65\textwidth}
        \centering
        \resizebox{1\textwidth}{!}{%
        \begin{tabular}{l | l | l}
        \textbf{Parameter}                                                                                       & \textbf{Formula}                                                           & \textbf{Value}                                \\
        \hline \hline
         
Memory cell size on EMCCD plane                                                                          & $1/\text{plate scale} \times \text{cell size} (0.0707^{\circ})$            & 2.32 mm                             \\ \hline
Total central width                                                                                      & $\lambda  \times \text{output f-number( = 51)}$, $\lambda$ = 0.487 $\mu$ m & 30.3 $\mu$m                         \\ \hline
Total FoV                                                                                                & plate scale $\times$ No of pixel $\times$ pixel size                       & 15.56 arcmin                        \\ \hline
\begin{tabular}[c]{@{}l@{}}Total number of cells across along one axis (within CCD \\ area)\end{tabular} & FoV/(size of each cell)                                                    & 3.67                                \\ \hline
Geometric encircled energy                                                                               & 60 $\mu$m @487 nm                                                          & 16 $\mu$m ; 3.75$\times$3.75 pixels  \\ \hline

        \end{tabular}}
        \caption{}
        \label{tab:EMCCD_1}
     \end{subtable}
     \caption{a) Important parameters encapsulating best collimated beam, b) Technical details of memory cells falling on the sensor }
     
\end{table}

\subsection{Tolerance analysis}
\label{sec:SIMULATOR_tolerance}

\begin{figure}[h!]
    \centerline{\includegraphics[width=0.33\textwidth]{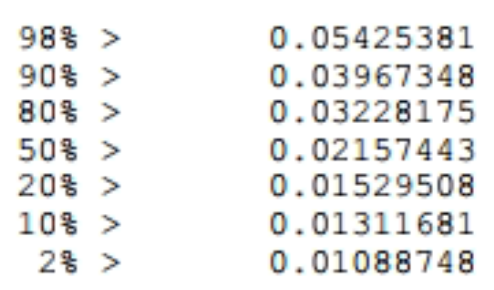}}
    \caption{\label{fig:monte_carlo} Monte carlo outcome from 2000 iterations using normal distribution. Diffraction limit spot size at 0.633 $\mu$m is 39.38 $\mu$m }
\end{figure}

Fig.~\ref{fig:monte_carlo} shows the outcome of Monte Carlo with iterations over 2000, using a normal distribution function. For an air disk radius of 40 $\mu$m at 0.633 $\mu$m, 90\% of the iterations are within an airy disk radius. This result corresponds to a confidence level of 1.5 sigmas (or corresponds to 87\% ). That means over 90\% of the time, the RMS radius stays within airy disc size.

\section{Optomechanical design}
\label{sec:opto_final_design}

Opto-mechanical design for SIMULATOR has been led by two core team members of IUCAA: Abhay and Bhushan. Abhay has expertise in SolidWorks, and Bhushan has PCB layout and design expertise. This section reviews the preliminary design and final opto-mechanical design of SIMULATOR.

\begin{figure}
    \centering
    \begin{subfigure}[b]{0.48\linewidth}        
        \centering
        \includegraphics[width=\linewidth]{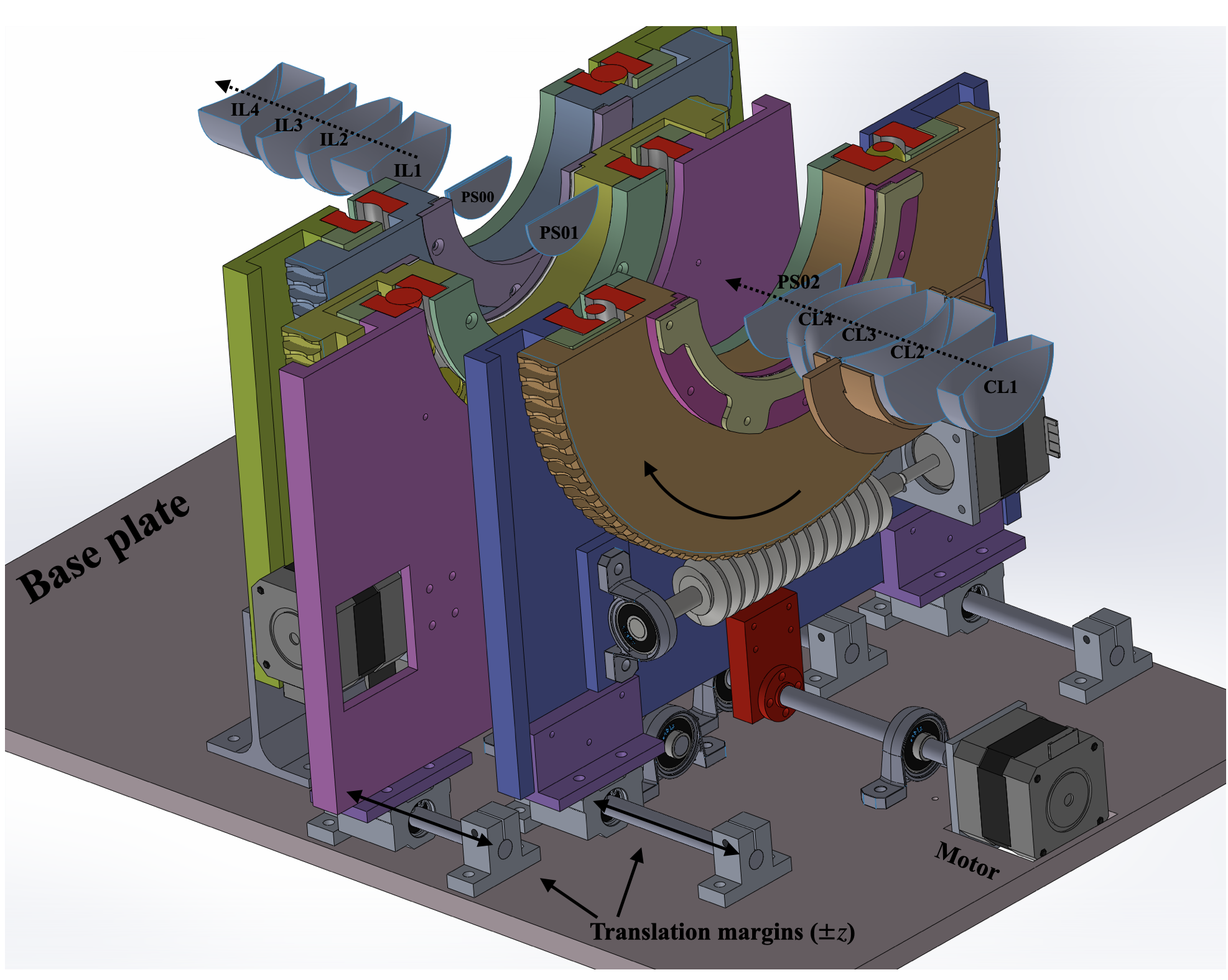}
        \caption{\label{fig:preliminary_phasescreen}}
    \end{subfigure}
    \begin{subfigure}[b]{0.48\linewidth}        
        \centering
        \includegraphics[width=\linewidth]{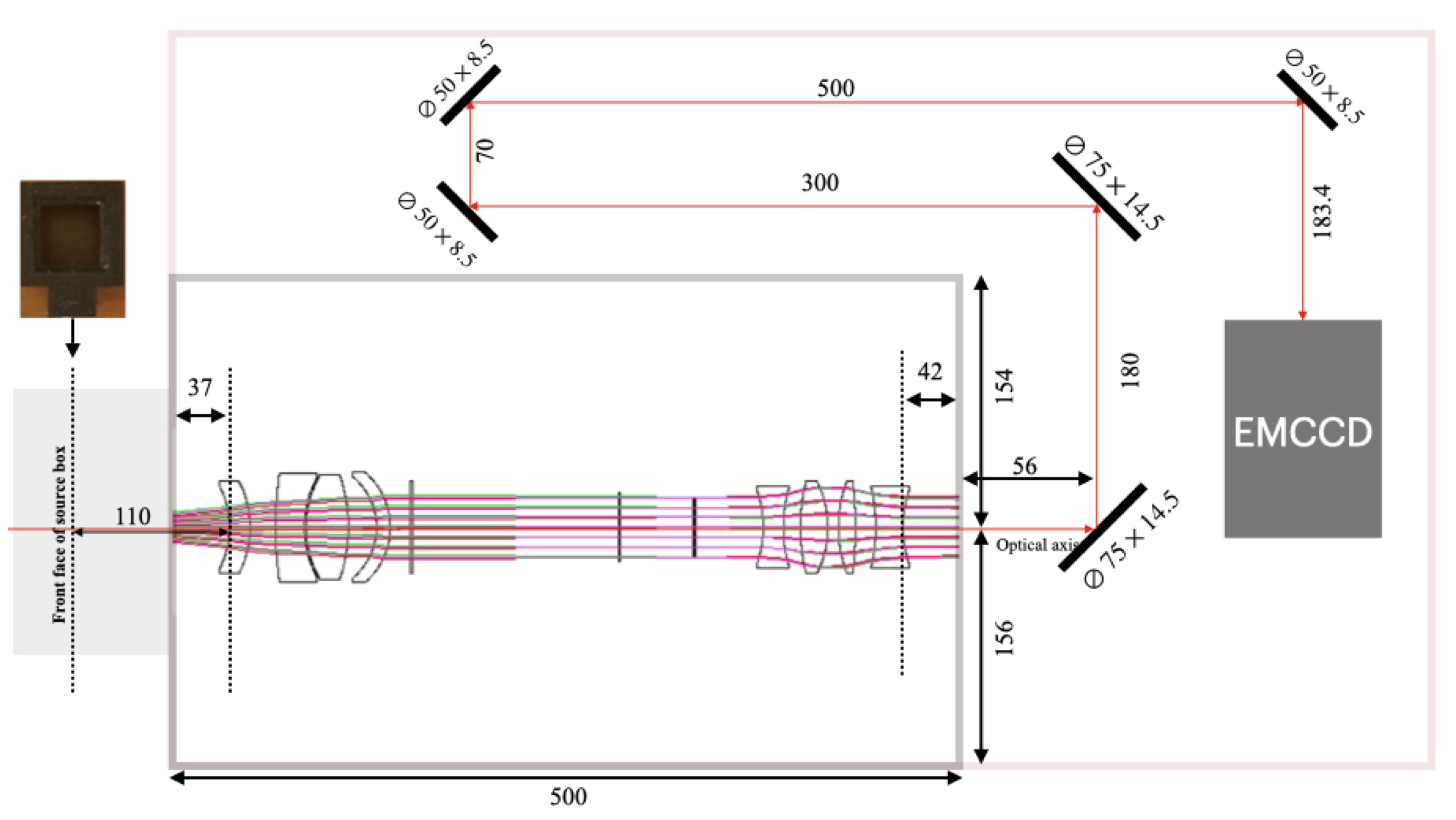}
        \caption{\label{fig:OM_mirror}}
    \end{subfigure}
    \caption{(a) Preliminary design of all three phase screens adjusted on base plate, along with rotation and translation stages, (b) Schematic layout for proposed mirror design for optimal space coverage}
    \label{fig:roc_curve}
\end{figure}

Fig.~\ref{fig:preliminary_phasescreen} shows a preliminary design for all three turbulence phase screens adjusted on a base plate. All three bearings have been procured from SKF with sizes and thicknesses relevant to each turbulence phase screen. All three stands for phase screens have a rotating mechanism at the bottom of each plate. Additionally, two linear translation stages are set at the top of the base plate for PS01 and PS02. Relevant motors with enough high torque to balance out the heavyweight of the entire structure have been procured from robokits.     

\begin{figure}[h!]
    \centerline{\includegraphics[angle=0,width=0.7\textwidth]{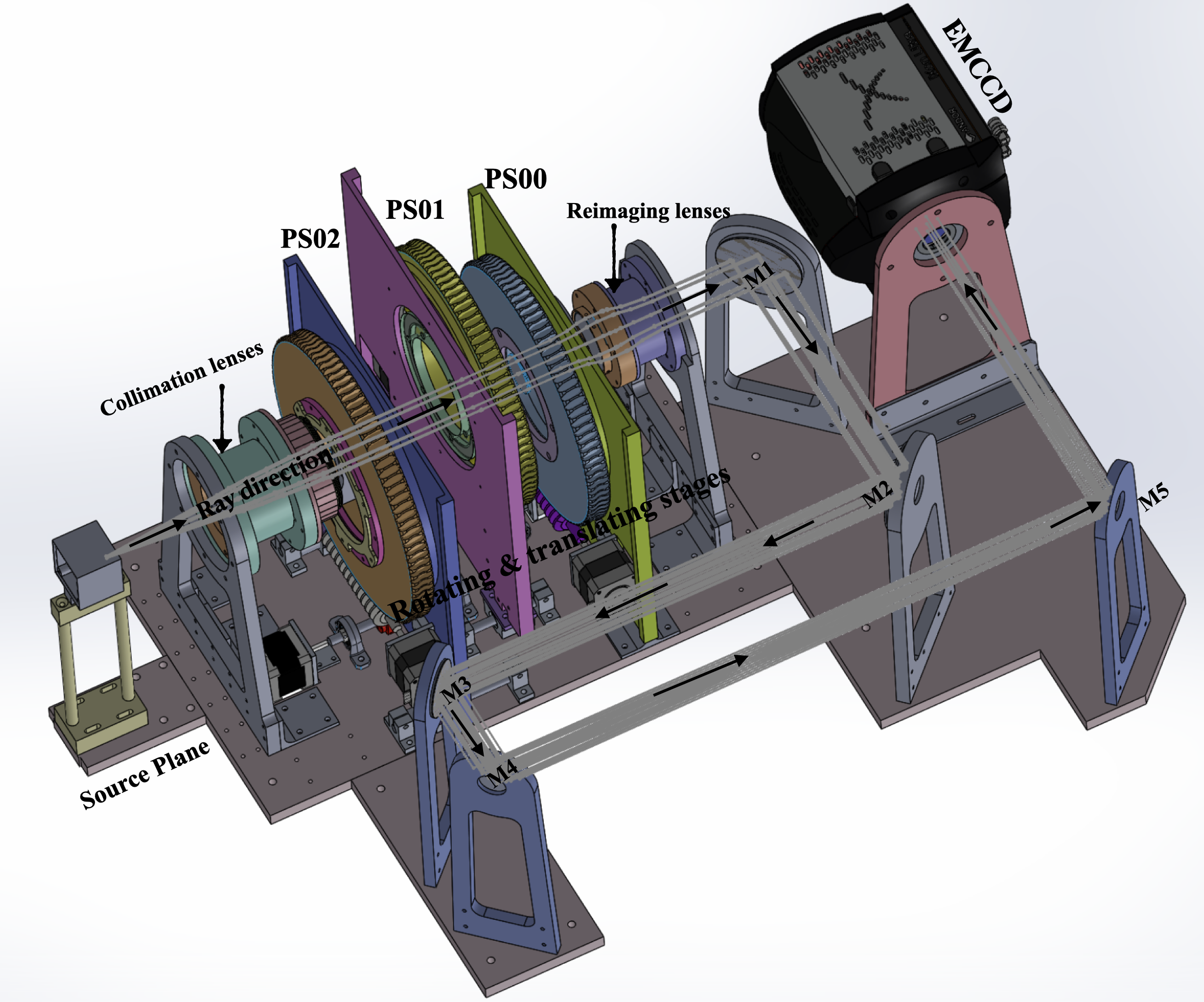}}
    \caption{\label{fig:OM_design} Optomechanical design of SIMULATOR }
\end{figure}

Fig.~\ref{fig:OM_design} shows the final optomechanical design of SIMULATOR. Starting from the source plane sitting on a stand, a ray of light first passes through a set of fixed collimation lenses (as one body system: Fixed CL1-3) and to the compensator CL4. After passing through three turbulent rotating phase screens, it will produce a collimated beam of pupil size 35.355 mm. The pupil plane has been placed within PS00 (not visible in this picture). The final pupil location has been adjusted (a little further away from PS00 due to space constraints), and thus accordingly, size has been adapted for the pupil plane. Adjustable lens IL1 (independent from the rest of the lenses) and three fixed lens systems (IL2-4) are aligned to produce a slow converging beam. Last, five-fold mirrors M1-5 have been deployed to make the system compact and finally to the EMCCD. EMCCD holder can withstand small translation in the y-axis and z-axis for readjustment.      

\section{Assembly, integration and testing (AIT)}
\label{sec:AIT}

\begin{figure}[h!]
    \centerline{\includegraphics[width=0.6\textwidth]{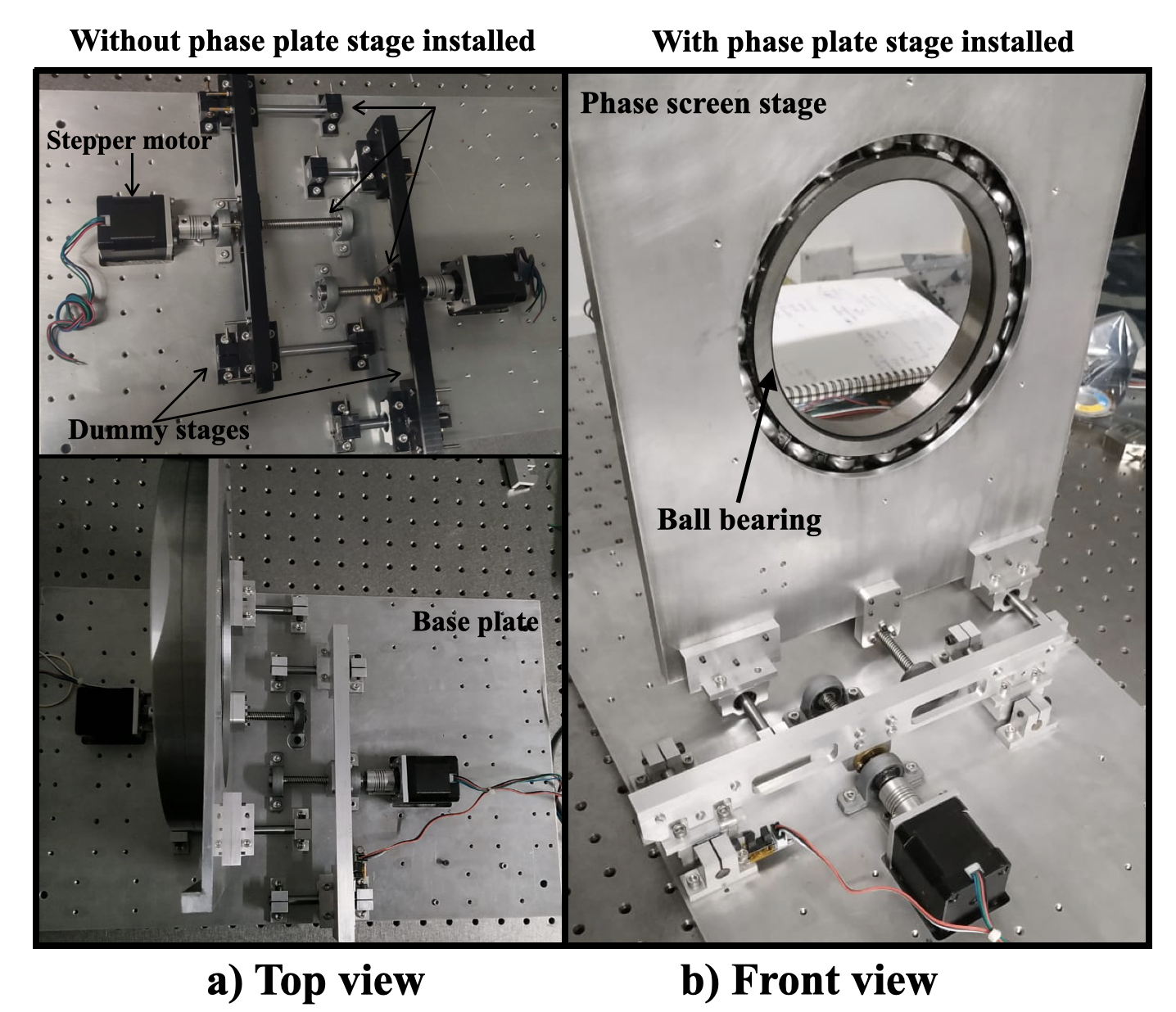}}
    \caption{\label{fig:stage_testing} Integration and testing of translation stages (both hardware and software) of three turbulent phase screen.}
\end{figure}

Fig.~\ref{fig:stage_testing} shows the setup for in-house made translation stages for three turbulent phase screens fitted over a flat stage. Motors can be read through an in-house hardware element called white dwarf (made by Bhushan and tested by Rani, instrumentation lab, IUCAA). This element will read user commands from the PC, act as an inter-mediator and pass signals in step voltages to the motor with the best accuracy. Hardware and software integration of the motors are already in place.

\subsection{Electronics design}
\label{sec:electronics_final_design}
\begin{figure}
    \centering
    \begin{subfigure}[b]{0.48\linewidth}        
        \centering
        \includegraphics[width=\linewidth]{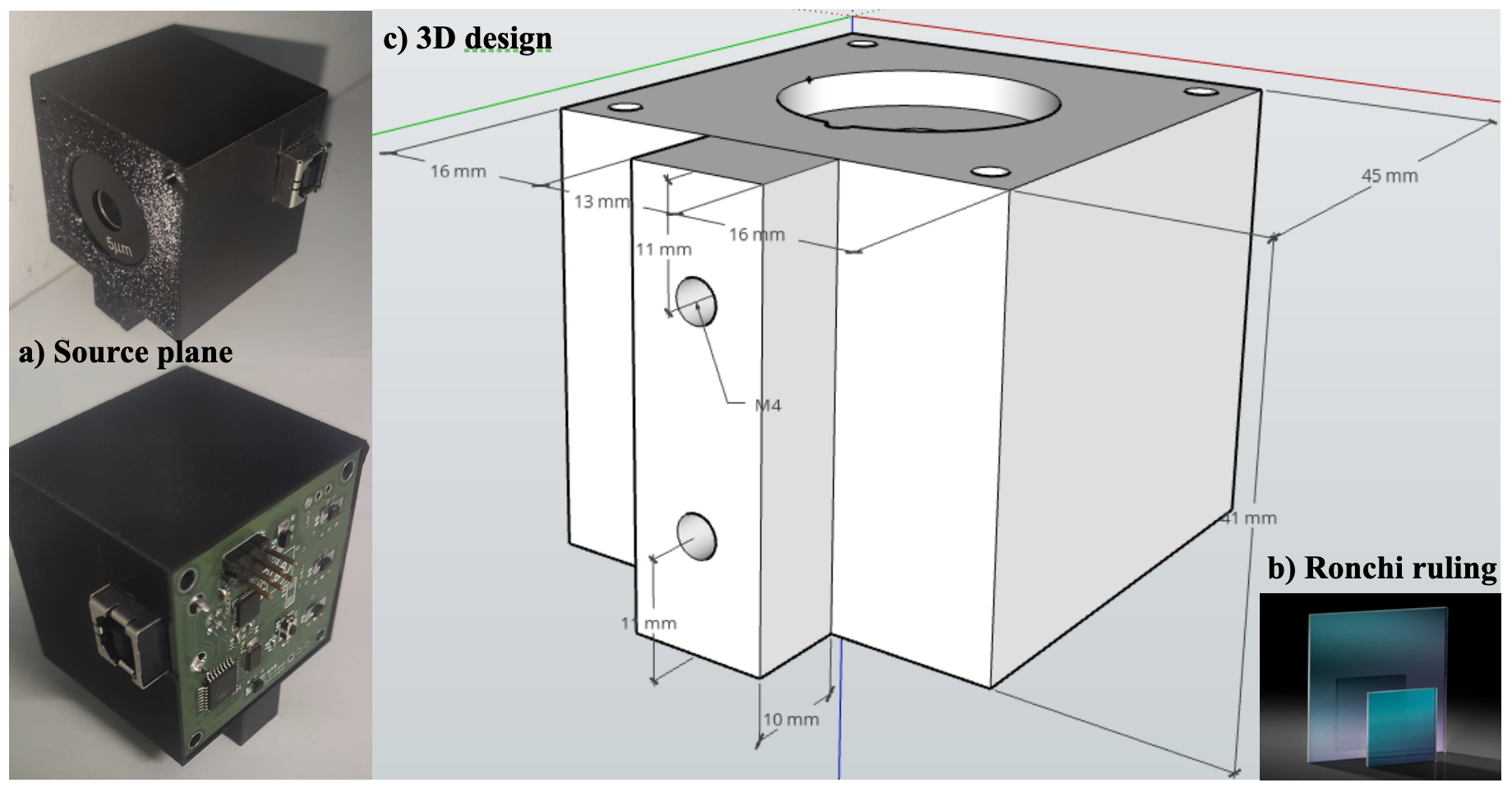}
        \caption{\label{fig:OM_source}a) 3D printed design of source plane box, b) Target: ronchi ruling 2000 lpi and c) Dimensions of source plane box}
    \end{subfigure}
    \begin{subfigure}[b]{0.48\linewidth}        
        \centering
        \includegraphics[width=\linewidth]{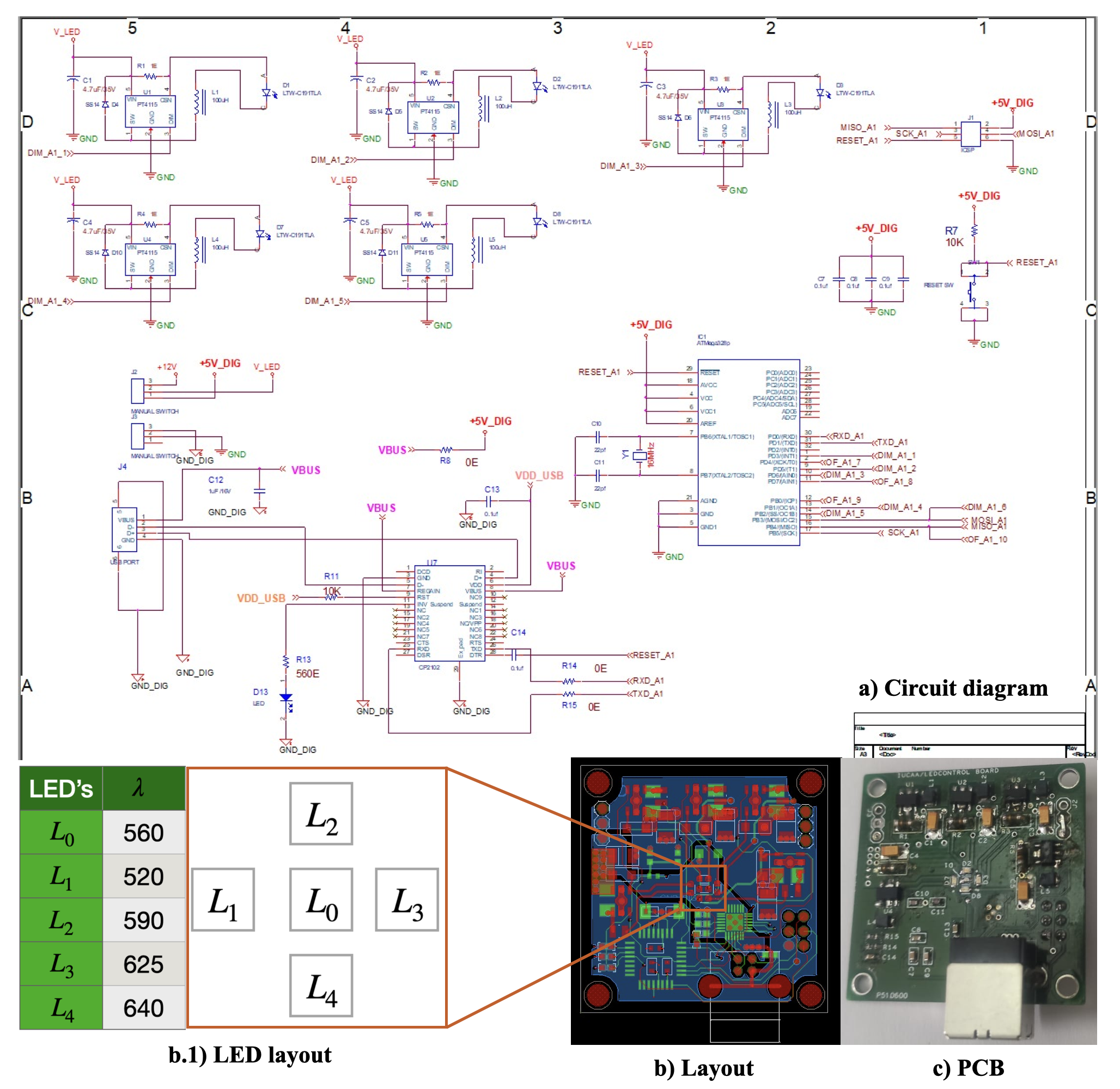}
        \caption{\label{fig:OM_PCB} a) Circuit diagram, b) Layout, b.1) LED layout and c) PCB design of LED's control board}
    \end{subfigure}
    \caption{Electronics circuit boards setup for SIMULATOR}
\end{figure}

Fig.~\ref{fig:OM_source} shows a final 3d printed design for source plane box. The box's front end is equipped with a 5-micron hole (as an example), and just behind that (inside the box), a target has been placed of 2000 lpi Ronchi ruling, which corresponds to 7 microns roughly. The back end of the box is equipped with a PCB board 
controlled by a USB connection to a computer. GUI has been designed through the MATLAB platform to control LEDs. Fig.~\ref{fig:OM_PCB} shows the circuit diagram layout for all 5 LED's covering wavelength bands from 520 - 640 nm. They have been aligned symmetrically with the central LED $L_0$ to affect the target ruling uniformly. For the experiment purpose, one LED will be used at a time. GUI for LED control has been produced and will be integrated with the rest of the instrument control (shown in fig.~\ref{fig:GUI}).

\subsection{Graphic user interface}
\label{sec:GUI}
\begin{figure}[h!]
    \centerline{\includegraphics[width=0.5\textwidth]{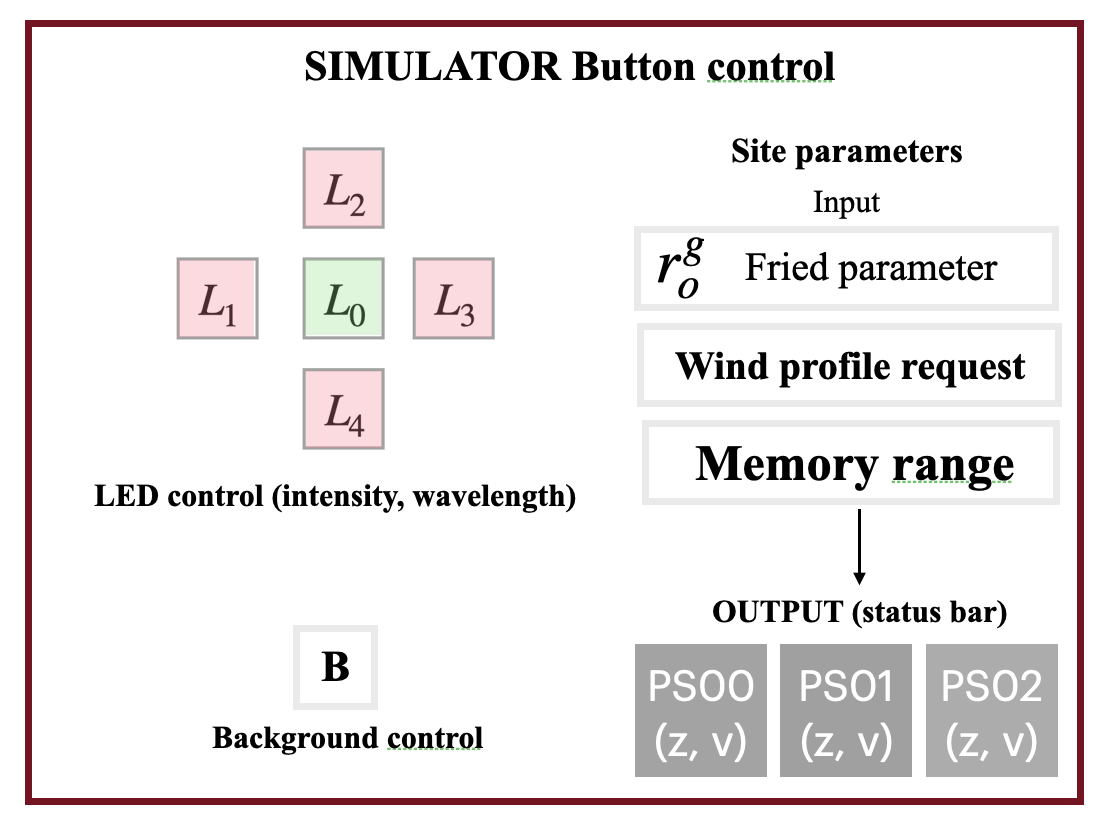}}
    \caption{\label{fig:GUI} Graphic user interface (GUI) window for complete automation of SIMULATOR }
\end{figure}

Fig.~\ref{fig:GUI} shows a user interfacing window for complete automation of SIMULATOR. The current version can control commands for LEDs, background luminosity and site parameters. However, work is still in progress, and the final design will also incorporate commands for controlling EMCCD.  

\acknowledgments 

\bibliography{report} 
\bibliographystyle{spiebib} 

\end{document}